\documentclass[%
 preprint,
superscriptaddress,
nofootinbib,
 amsmath,amssymb,
 aps, physrev,
]{revtex4-2}

\usepackage{graphicx}
\usepackage{dcolumn}
\usepackage{bm}
\usepackage{subfiles}
\usepackage{import}

\usepackage{comment}
\usepackage{float}
\usepackage{upgreek}
\usepackage{hyperref}
\usepackage[mathlines]{lineno}
\usepackage{siunitx}
\usepackage{xcolor}
\usepackage{setspace}
\usepackage{physics}
\usepackage{soul}

\newcommand{\tn}[1]{\mathrm{#1}}
\newcommand{\vch}{$V_{\tn{ch}}$}
\newcommand{\vgt}{$V_{\tn{gt}}$}
\newcommand{\vres}{$V_{\tn{res}}$}

\newcommand{\vac}{$V_{\tn{ac}}$}
\newcommand{\fac}{$f_{\tn{ac}}$}

\begin{document}

\title{\textbf{Plasmon mode engineering with electrons on helium}
}%

\author{Camille~A.~Mikolas}
\email{mikolasc@msu.edu}
\affiliation{Department of Physics \& Astronomy, Michigan State University, East Lansing, MI 48824, USA}

\author{Niyaz~R.~Beysengulov}
\affiliation{Department of Physics \& Astronomy, Michigan State University, East Lansing, MI 48824, USA}
\affiliation{EeroQ Corporation, Chicago, IL 60651, USA}

\author{Austin~J.~Schleusner}
\affiliation{Department of Physics \& Astronomy, Michigan State University, East Lansing, MI 48824, USA}

\author{David~G.~Rees}
\affiliation{EeroQ Corporation, Chicago, IL 60651, USA}

\author{Camryn~Undershute}
\affiliation{Department of Physics \& Astronomy, Michigan State University, East Lansing, MI 48824, USA}

\author{Johannes~Pollanen}
\email{pollanen@msu.edu}
\affiliation{Department of Physics \& Astronomy, Michigan State University, East Lansing, MI 48824, USA}


\maketitle

\noindent \textbf{An ensemble of electrons trapped above superfluid helium offers a paradigm system for investigating and controlling collective charge dynamics in low-dimensional electronic matter. Of particular interest is the ability to spatially control and engineer surface plasmons for integration with hybrid quantum systems and circuit quantum electrodynamic device architectures. Here we present experiments using an electron-on-helium microchannel device that hosts microwave-frequency plasmons, generated via local microwave excitation in an electrostatically defined central channel. By precisely varying the electron density, we demonstrate tunability of plasmon mode frequencies over several GHz. Additionally, we find that the power dependence of these modes can be used to investigate both homogeneous and inhomogeneous sources of spectral broadening. These results demonstrate the versatility of electrons on helium for probing collective excitations in low-dimensional Coulomb liquids and solids, and demonstrate a path for integrating engineered plasmons in electrons on helium with hybrid circuit quantum electrodynamic systems.}

\vspace{0.6cm}

\noindent \textbf{\large{Introduction}}

\noindent Circuit quantum electrodynamics (cQED)~\cite{Blais2021} has enabled the development of sophisticated quantum control and measurement protocols for a wide variety of quantum systems ranging from superconducting circuits~\cite{Blais2004} and semiconductor spins~\cite{Petersson2012}, to systems of trapped electrons~\cite{koolstra2019coupling, zhou2022single, Zhou2023}, as well as nano- and micromechanical oscillators~\cite{lahaye2009nanomechanical,o2010quantum, Pirkkalainen2013}. These techniques can also be leveraged as powerful experimental tools for investigating microwave frequency collective phenomena in quantum systems composed of many interacting particles or degrees of freedom~\cite{clerk2020hybrid}. For example, when coupled with superconducting circuits, these approaches have been used to study collective modes in magnonic~\cite{goryachev2014high, tabuchi2015coherent, li2018magnon, lachance2020entanglement} and phononic~\cite{gustafsson2014propagating, chu2017quantum, satzinger2018quantum, kitzman2023phononic} systems and to investigate the dynamics of spin ensembles~\cite{kubo2010strong, schuster2010high}.

Electrons trapped above the surface of condensed noble gas substrates, such as superfluid helium or solid neon, are emerging as promising systems for integration with cQED architectures and microwave frequency devices for quantum information processing~\cite{platzman1999quantum, dykman2023spin, lyon2006spin, Beysengulov2024}. At the level of single electrons, cQED techniques have been used to investigate the in-plane orbital states of electrons on helium~\cite{schuster2010proposal, koolstra2019coupling} and have recently been utilized to realize high-coherence charge qubits on the surface of solidified neon~\cite{zhou2022single, Zhou2023}. In contrast to single electron dynamics, these systems can also host a wide variety of collective charge modes including plasmonic~\cite{grimesadams, grimes1979evidence} and magneto-plasmonic excitations~\cite{mast1985observation,Glattli1985,Lea1994,Chepelianskii2021,kostylev2024delocalized}, as well as hybrid modes coupling the dynamics of multiple degrees of freedom~\cite{yunusova2019coupling, zadorozhko2021motional,byeon2021piezoacoustics}. Additionally, ensembles of electrons on helium have been strongly coupled to three-dimensional microwave cavities to study cyclotron resonance~\cite{abdurakhimov2016strong, chen2018strong} and integrated into hybrid circuits in which an electron ensemble is placed above a planar microwave resonator~\cite{yang2016coupling}.
Fully leveraging cQED-type techniques to study the high-frequency dynamics of electrons on helium requires the development of devices that have not only an optimized microwave environment~\cite{yang2016coupling}, but also the ability to engineer and manipulate the collective modes of the electron system via precise spatial control. In this work, we address the latter of these aspects by realizing a device that enables precision control over the spatial distribution of electrons in a microchannel geometry, providing the ability to engineer, excite, and detect plasmonic excitations with frequencies in a range compatible with cQED-based systems. Local microwave excitation resonantly couples to the plasmon modes, which we detect via changes in the electron conductance determined by simultaneous transport measurements. By precisely varying the electron density in the microchannel, we can tune the frequency of the modes by several GHz. Analyzing the power dependent plasmon response allows us to investigate possible mechanisms leading to plasmon dephasing and energy loss. Finally, we highlight how this type of device and our results demonstrate the overall system control necessary to integrate with future low-loss microwave cQED architectures.

\vspace{0.5cm}

\noindent \textbf{\large{Results and discussion}}

\noindent \textbf{Microchannel device for plasmon confinement}

\noindent Microchannel device architectures, like the one we employ, are widely used to study the effect of geometric confinement on the thermodynamic ground state and transport properties of electrons on helium. Typically in these devices, micron-scale deep channels are filled with superfluid helium via capillary forces and electrons are deposited above the superfluid surface. Metallic electrodes around the channels are used to precisely shape the electrostatic environment to control the spatial distribution of surface state electrons and perform transport experiments~\cite{marty1986stability,rees2011point}. These types of channeled devices have been used to reveal dynamical ordering of two-dimensional~\cite{rees2016stick,ZouPRB2021} and quasi-one-dimensional electron chains~\cite{glasson2001observation, rees2016structural}, and perform ultra-efficient clocking of electrons in microchannel-based CCD arrays~\cite{sabouret2008signal, bradbury2011efficient}. Here we leverage a microchannel architecture to engineer the spatial structure of the two-dimensional electron system in order to host and investigate charge density oscillations, i.e. plasmons.

\begin{figure}[h]
    \centering
    \includegraphics[width=1\textwidth]{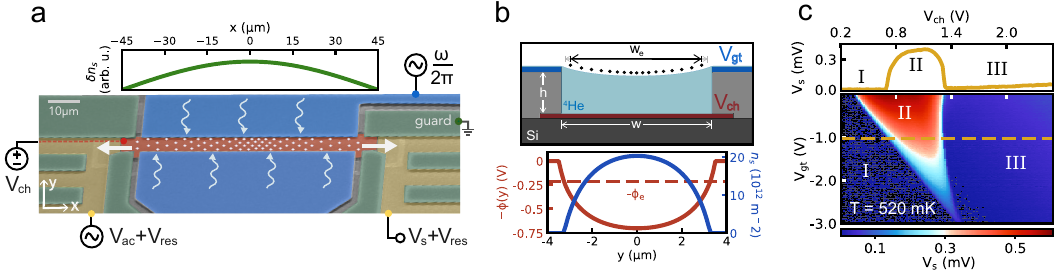}
    \caption{\textbf{Microchannel device for plasmon-mode engineering}. (a) False color scanning electron micrograph of the microchannel device. Grounded guard electrodes (green) patterned above a resist layer surround reservoir electrodes (yellow) located beneath. The central microchannel region between the two reservoirs consists of a channel electrode (red) and top side gate electrode (blue). An ac voltage $V_{\mathrm{ac}}$ at frequency $f_{\mathrm{ac}}$ drives the electrons (white dots) through the channel and the resulting transport signal $V_{\mathrm{s}}$ is detected via a lock-in amplifier. A microwave signal with a frequency $\omega/2\pi$ is applied to the gate electrode to generate longitudinal plasmons in the central channel (the first mode of which is shown schematically by the $\delta n_s$ plot). (b) Top panel: Schematic cross section view across ${x=0}$ of the channel with a fixed dc bias voltage (i.e. no microwave drive) on the side gate electrode $V_{\mathrm{gt}}$. Bottom panel: Electrostatic potential profile, $\phi(x=0,y)$, transverse to the channel (solid red line), with chemical potential $\phi_e$ (dashed red line), and the distribution of electron density $n_s(y)$ along the $y$-direction (solid blue line). (c) Transport measurements performed at various values of $V_{\mathrm{gt}}$, showing the three characteristic transport regimes. Here, $V_{\mathrm{res}} = 0.9$~V, $V_{\mathrm{ac}} = 20$~mV, $f_{\mathrm{ac}} = 1.408$~MHz. See main text for complete description.}
    \label{fig:dev1}
\end{figure}

The device is fabricated on a 7~mm $\times$ 2~mm high resistivity silicon chip, onto which hard-baked resist is deposited and selectively etched to create $h\simeq 1.4~\upmu$m deep channels. As shown in Fig.~\ref{fig:dev1}a, four electrodes are lithographically patterned to define two reservoir areas connected via a central microchannel region having a length of $L=90~\upmu$m and a width of $w=7~\upmu$m. The device is placed into a superfluid-leak-tight sample cell that is mounted onto the mixing chamber plate of a dilution refrigerator. Helium gas from a room temperature volume is introduced into the sample cell, where it condenses into a superfluid and fills the channels via capillary action. Electrons are deposited onto the superfluid surface via thermal emission from a tungsten filament. 

The high degree of spatial control and confinement over the electrons in the reservoirs and microchannel regions are enabled by voltages ($V_i$) applied to the four electrodes: $V_{\mathrm{gt}}$~(gate), $V_{\mathrm{ch}}$~(channel), $V_{\mathrm{res}}$~(reservoirs), and the guard electrode. These voltages allow us to control the two-dimensional electrostatic environment experienced by the electrons in the plane of the helium surface $\phi(x, y) = \sum_i V_i \alpha_i(x, y)$ (see~Fig.~\ref{fig:dev1}b), where the constant $\alpha_i(x,y)$ describes the capacitive coupling between the electrons and the corresponding $i^{\mathrm{th}}$ electrode. To design a given confinement profile, we numerically solve the Laplace equation using finite element modeling (FEM) techniques~\cite{beysengulov2016structural}. This allows us to extract $\alpha_i$ and construct the potential by applying appropriate values of $V_i$. This numerical procedure also allows us to calculate the areal electron density $n_s(x,y)$ for a given potential. 

By controlling the electrostatic environment in this fashion, we can effectively create a resonant cavity for confined plasmonic modes, i.e. oscillations of the charge density $\delta n_s$ along the length of the central channel. At the boundaries of the microchannel, electrons are free to enter and exit into the reservoir regions, enforcing charge density nodes at the ends of the channel, as depicted in Fig.~\ref{fig:dev1}a. Because the number of electron rows is $\gtrsim 10$ in the density regime in which we investigate plasmons, the electrons in the central channel can be modeled as a two-dimensional sheet of charge defined by the electrostatic confinement produced by the electrode voltages. The long and narrow geometry ($L\gg w$) of the central channel ensures a large separation in frequency between plasmons along versus perpendicular to the channel. This allows us to consider only longitudinal plasmon standing-waves along the channel length, which have the following dispersion relation~\cite{grimesadams, monarkha2004two, andrei1997two}, 
\begin{equation}
    \omega_p^2 = \frac{n_s e^2}{2 \varepsilon_0 m_e} \sqrt{q_x^2 F(q_x)}~,
    \label{eq:plasmon}
\end{equation}
\noindent where $\omega_p$ is the density-dependent frequency of a plasmon having wavevector $q_x = n \pi / L$ and mode number $n$, $m_e$ is the electron mass, $e$ is the electron charge, and $\varepsilon_0$ is the vacuum permittivity (see Supplementary Information, section I). The wavevector-dependent factor $F(q_x)$ takes into account the reduction in the electron-electron interaction due to the presence of the nearby metallic electrodes. This screening factor will correspondingly reduce the plasmon frequency~\cite{grimesadams}, and for the geometry of our device we utilize the following phenomenological form for $F(q_x)$,
\begin{equation}
    F(q_x) = \frac{1}{2}(\tanh{q_x l} + \tanh{q_x h}),
    \label{eq:screen}
\end{equation}
\noindent where $l = w - w_e$ parameterizes the effective distance of the electron sheet from the surrounding side gate electrodes and $h = 1.4~\upmu$m is the height of the electrons above the bottom channel electrode. The effective width $w_e$ of the electron system is defined where the parabolic confinement potential of the channel is equal to the chemical potential $\phi_e$, i.e. $\phi(y=w_e/2) = \phi_e$, as shown in the bottom panel of Fig.~\ref{fig:dev1}b, which we extract from FEM. For our specific device geometry, which includes laterally defined side gate electrodes, an analytical solution for the screening factor is lacking. However, we find that the phenomenological form presented in Equation~(\ref{eq:screen}) captures to good approximation the screening contributions in the long wavelength limit ($q_xh, \ q_xl \ll 1$), as well as in the limiting case, in which the screening electrodes are moved infinitely far from the electrons in the channel and the unscreened plasmon dispersion is recovered, i.e. $F(q_x) = 1$. Using Equation~(\ref{eq:plasmon}) and FEM calculations, we find a fundamental ($n=1$) plasmon mode frequency of $\omega_{p}/2 \pi \simeq 1.0$~GHz at $n_{s} \simeq 2.3 \times 10^{12}$~m$^{-2}$. This lowest frequency mode corresponds to a half-wavelength standing wave of the time-varying change in density $\delta n_s$ along the channel, as shown in the top panel of Fig.~\ref{fig:dev1}a. In the following section, we discuss how these modes are generated using an additional microwave drive and detected using transport techniques. \\

\noindent \textbf{Transport measurements \& microwave excitation}

\noindent To characterize the electron system in the central microchannel, and its collective dynamics, we utilize a conventional ac transport measurement scheme~\cite{glasson2001observation}. In these measurements, an ac voltage $V_{\mathrm{ac}}$  is superimposed on the left reservoir electrode driving electrons from one reservoir to the other via the central channel at a frequency $f_{\mathrm{ac}}$. The resulting electron transport through the channel is detected from the voltage $V_\textrm{s}$  induced on the right reservoir electrode, which we measure using standard phase-sensitive lock-in techniques. As described above, a dc voltage $V_{\mathrm{ch}}$ applied to the channel electrode controls the population of electrons in the central microchannel. In Fig.~\ref{fig:dev1}c we show a standard transport map as we tune the electron density and confinement potential in the central channel. This type of measurement reveals three transport regimes depending on the density of electrons in the central microchannel. In regime I, $\phi(y=0) < \phi_e$, and electrons cannot enter the channel from the reservoirs. When $\phi(y=0) = \phi_e$, the channel threshold voltage $V_{\mathrm{ch}}^{\mathrm{th}}$ condition is met and electrons can enter the microchannel for $V_{\mathrm{ch}} \geq V_{\mathrm{ch}}^{\mathrm{th}}$. In this regime (regime II), the electrons form a highly conducting state in which the electrons interact weakly with the helium surface, resulting in a large transport signal~\cite{rees2016stick}. At sufficiently high density, the electrons in the microchannel form a low-conductivity Wigner solid (regime III)~\cite{dykman1997bragg}. These measurements also allow us to calculate the electron density in the central microchannel $n_s$ from the potential in the center of the channel $\phi^0 \equiv \phi (x = 0, y = 0)$ and the chemical potential $\phi_e$, as
\begin{equation}
    n_s = \frac{\varepsilon_{\mathrm{He}} \varepsilon_0}{e h} (\phi^0 - \phi_e),
    \label{eq:dens}
\end{equation}
\noindent where $\varepsilon_{\mathrm{He}} = 1.057$ is the dielectric constant of liquid helium. Here, the chemical potential is calculated using ${\phi_e = V^{\mathrm{th}}_{\mathrm{ch}}\alpha^0_{\mathrm{ch}}}$~\cite{rees2012} and the capacitive coupling constant in the center of the microchannel, ${\alpha^0_{\mathrm{ch}} \equiv \alpha_{\mathrm{ch}} (x=0,y=0) \simeq 0.7}$, is obtained through FEM calculations of the device~\cite{rees2016structural}. 

To generate plasma oscillations in the electron sheet, we apply a high-frequency signal onto the gate electrodes located on either side of the microchannel as shown in Fig.~\ref{fig:dev1}a. The microwave power modulates the otherwise static confinement potential throughout the microchannel, which leads to a periodic modulation of the effective width $w_e$ of the electron system (see top panel Fig.~\ref{fig:dev1}b) and creates charge density oscillations of the electrons in the central microchannel due to the strong Coulomb interaction. The frequency and amplitude of these oscillations are controlled by the gate modulation frequency $\omega/2 \pi$ and microwave signal power $P$, which is measured from the output of the high-frequency source. This microwave signal is attenuated by an additional 28~dB before entering the cryostat.

\begin{figure}[h]
    \centering
    \includegraphics{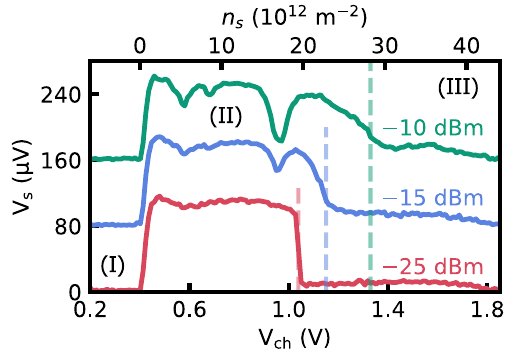}
    \caption{\textbf{Transport detected plasmon resonances.} Microchannel transport measurements in the presence of a $\omega/2\pi~=~5.5$~GHz microwave excitation signal on the gate electrode for increasing values of microwave power $P$. Traces are offset vertically for clarity and color coded with the corresponding $P$. Vertical dashed lines indicate the transition into a low-conductivity Wigner crystal electron state, which increases with increasing $P$. At $P \gtrsim -15$~dBm, resonance-like features appear in regime II, indicating plasmon modes generated along the length of the channel as described in the main text. Measurements were performed at $T=18$~mK, with $V_{\mathrm{gt}}=0$~V, $V_{\mathrm{ac}}=6$~mV, $f_{\mathrm{ac}}=3$~MHz, and $V_{\mathrm{res}}~=~0.4$~V.}
    \label{fig:heat}
\end{figure}

\vspace{0.5cm}

\noindent \textbf{Plasmon detection}

\noindent In Fig.~\ref{fig:heat}, we show how the GHz-frequency plasmonic modes of the electrons in the central microchannel are imparted on the transport signal by monitoring $V_\mathrm{s}$ as a function of $V_{\mathrm{ch}}$ as we increase the power $P$ of a fixed $\omega/2\pi~=~5.5$~GHz signal applied to the gate electrodes. In these measurements, we observe the three characteristic transport regimes described in the previous section, corresponding to (I) no electrons, (II) a low-density, highly conducting electron state, and (III) a high-density, low-conductivity Wigner solid within the central channel. At increasing microwave power, the electron density $n_s$ that corresponds to the transition into the low-conductivity Wigner solid regime increases~\cite{rees2020dynamical}, as shown by the vertical dashed lines for each trace, and we find that this type of effect occurs independent of the microwave drive frequency.

More interestingly, with increasing microwave power, we observe the emergence of resonance-like features in the transport signal when the electrons in the microchannel are in the highly conducting state (regime II). Three resonances are clearly visible in the green trace in Fig.~\ref{fig:heat} at $V_{\mathrm{ch}}=0.58,~0.68,~0.96$~V corresponding to densities of ${n_s~=~5.7,~8.6,~16.9~\times~10^{12}~\mathrm{m}^{-2}}$ in the central channel. The resonances appear as local minima in the transport signal indicative of a reduction in the conductivity of the electron system in the central channel.

To understand these experimental features, we must consider the non-linear transport phenomena arising from the coupling of the electrons to the helium surface when the system is subjected to the effects of the ac driving and microwave excitation fields, which drive the electrons out of equilibrium~\cite{konstantinov2008microwave,nasyedkin2011transport,zou2022observation}. We begin by noting that, at $T\simeq20$~mK, in the absence of microwave excitation and at low ac drive, the electron system would remain in the low-conductivity regime (III) for all values of the electron density shown in Fig.~\ref{fig:heat}. In other words, the equilibrium state of the electrons in this case would correspond to the low-conductivity Wigner solid\footnote{In this regime the electron system is in equilibrium with the helium bath, thus for an electron temperature $T_e~=~20$~mK the critical density to form a Wigner solid ($n_s^{\mathrm{cr}}~=~8~\times~10^9$~$\mathrm{m}^{-2}$) is reached for ($V_{\mathrm{ch}}~-~V_{\mathrm{ch}}^{\mathrm{th}})~>~0.2$~mV.}. In this regime, the electrons coherently emit ripplons whose wavevectors match the reciprocal lattice vectors of the electron crystal. This phenomenon is known as the resonant Bragg-Cherenkov scattering effect~\cite{dykman1997bragg}. This effect results in an increased frictional force on the electrons and a saturation of their velocity at the phase velocity of the emitted ripplons $v_{\mathrm{ph}}~=~\omega_{\mathrm{r}}(G_1)/G_1$, where $\omega_{\mathrm{r}}(k)~=~\sqrt{\sigma_t/\rho\cdot~k^3}$ is the ripplon dispersion relation, $\sigma_t~=~358$~$\mu$N/m is the liquid helium surface tension, $\rho~=~145$~kg/m$^3$ is the liquid helium density, and $G_1~=~(8\pi^2n_s/\sqrt{3})^{1/2}$ is the first reciprocal lattice vector of the Wigner solid. When subjected to high driving fields, the electron system can heat~\cite{saitoh1977warm} and transition to a non-equilibrium state which has a high conductivity (regime II). This state has been interpreted as the formation of either a disordered electron liquid state~\cite{giannetta1991, konstantinov2008microwave} (melting model) or as a depinning transition of the Wigner solid~\cite{shirahama1995dynamical,ikegami2009nonlinear,rees2020dynamical,rees2016stick} (sliding model). In the sliding model, the transition to the Wigner solid state is associated with the formation of static surface deformations appearing under each electron, referred to as a dimple lattice, which moves together with the electron lattice. When subjected to a sufficiently strong drive, the electron solid can decouple from the underlying dimple lattice and move at higher velocities. In contrast, the melting model does not involve the concept of a dimple lattice; instead, the transition to a high-conductivity regime is interpreted as a transition to a disordered state. Despite the lack of an unambiguous microscopic description of these nonlinear effects, the transition into the high-conductivity regime (II) strongly depends on the positional order of the electrons in both models. Reducing this order weakens the Bragg-Cherenkov scattering effects and lowers the critical ac driving field required to transition into the high-conductivity state. For the experimental data presented in Fig.~\ref{fig:heat}, transport measurements were performed at ac driving fields sufficiently high enough to promote the electrons into the non-equilibrium high-conductivity state for densities $n_s<18.8~\times~10^{12}$~m$^{-2}$.

\begin{figure}[h]
    \centering
    \includegraphics[width=0.65\textwidth]{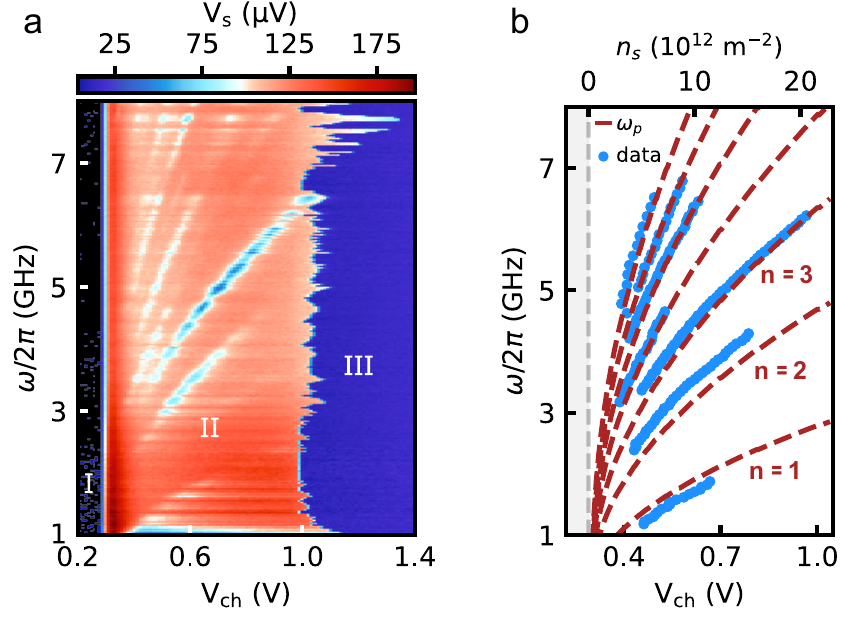}
    \caption{\textbf{Characterization and tuning of plasmon modes.} (a) Microwave frequency dependent transport map. Transport measurements are performed by sweeping $V_{\mathrm{ch}}$  to control the electron density in the central channel while simultaneously applying a microwave signal $\omega/2\pi$ to the gate electrodes. For $V_{\mathrm{ch}}~\lesssim~0.29~$V, the channel is empty (I). Above $V_{\mathrm{ch}}^{\mathrm{th}}$ electrons can enter the microchannel and form a low-density high-conductivity state (II) and high-density low conductivity Wigner solid (III). The transport signal $V_{\mathrm{s}}$ reveals a family of density and frequency dependent plasmon resonances that manifest as local minima in regime II. Measurements were performed at $P=-15$~dBm, with $V_{\mathrm{gt}}~=~0$~V, $V_{\mathrm{ac}}~=~8$~mV, $f_{\mathrm{ac}}~=~3$~MHz, $V_{\mathrm{res}}~=~0.29$~V, and $T=26$~mK. (b) Blue dots are the extracted local minima from the first seven resonances in (a). Red dashed lines correspond to the first seven longitudinal plasmon modes along the channel, calculated using the dispersion relation given in Equation~(\ref{eq:plasmon}) with our device design parameters. Grey shaded region indicates the empty channel (I) and grey dashed line indicates $V_{\mathrm{ch}}^{\mathrm{th}}~=~0.29$~V.}
    \label{fig:plas}
\end{figure}

The application of the additional microwave excitation field onto the side gate electrodes further perturbs the electron system. In the Wigner solid state, these perturbations induce high-frequency electron motion, which in a quasi-static approximation ($\omega\gg\omega_{\mathrm{r}}$) can be viewed as a weakening of the positional order of the electron solid. As a result, the transition into the low-conductivity Bragg-Cherenkov scattering regime shifts to higher in densities with increasing amplitude of the perturbing field. This effect is analogous to raising the temperature of the electron system $T_e$, which characterizes the melting of the solid.
Due to the large electron-electron collision rate (10$^{11}$~s$^{-1}$) and small energy relaxation rate ({$10^{5}-10^{6}~\mathrm{s}^{-1}$}), the electron system temperature can be raised above that of the helium bath~\cite{glattli1988thermodynamic}. In this way, the perturbing field effectively melts the Wigner solid leading to a transition into the high-conductivity regime. The ultimate electron temperature produced by the microwave field is determined by a balancing of the incident microwave and ac drive field powers with the energy transferred into the helium bath via the emission of short-wavelength ripplons and phonons~\cite{dykman2003, rees2020dynamical}. Due to the complex geometry of the device, which includes multiple regions with varying electron density, and the lack of information about how much of the incident microwave power is absorbed by the electron system, estimating $T_e$ is unfeasible. Nonetheless, a qualitative approach can be employed to interpret the decrease in the measured transport signal at the position of the resonances shown in Fig.~\ref{fig:heat}. 

Independent of the underlying microscopic state of the electron system in regime II, the microwave energy absorbed by the electrons increases when plasmons are resonantly excited, which results in additional heating of the electron system on resonance. Since the resonances appear in the high-conductivity regime (II), an additional increase in $T_e$ can be understood as producing an increase in the electrons' momentum transfer rate to ripplons, leading to a reduction in the mobility of the electron liquid state~\cite{saitoh1977warm}. This is consistent with the observed decrease in the measured signal on resonance. We note that, in principle, the reduction in the measured transport signal on resonance could also be interpreted as a transition from a unpinned Wigner solid into the Bragg-Cherenkov non-linear regime. However, our experiments indicate that in the presence of microwave excitation, the high-conductivity regime consistently remains in a linear transport regime, indicative of an electron liquid state (see Supplementary Information, section II).

These transport measurements, conducted with an additional perturbing microwave field, reveal the high sensitivity of the measured signal to the presence of plasmonic excitations in the electron system confined within the microchannel. These experiments underscore the complex nature of the non-equilibrium and nonlinear response of this strongly correlated low-dimensional electron system coupled to the helium surface excitations and enrich the extensive body of research on these topics~\cite{saitoh1977warm, konstantinov2008microwave, shirahama1995dynamical, ikegami2009nonlinear, nasyedkin2011transport}. Despite the absence of an unequivocal microscopic picture of the electron conductivity in these regimes, we can leverage the sensitivity of these measurements to investigate the plasmonic excitations we generate in the central microchannel. Finally, we note a similar technique has recently been employed to detect the excitation of Rydberg-like resonances of electrons on helium due to resonant microwave heating~\cite{zou2022observation}.

\vspace{0.5cm}

\noindent \textbf{Analysis of plasmon mode structure}

\noindent In Fig.~\ref{fig:plas}a we show the full channel density and microwave frequency dependence of the transport signal through the device. In regime II, we observe a family of density-dependent resonances in the channel, which are consistent with the long-wavelength two-dimensional longitudinal plasmons described by Equation~(\ref{eq:plasmon}) (see Supplementary Information, section~I). To analyze these plasmon modes, we extract the local transport minima along each of the first seven resonances (blue dots in Fig.~\ref{fig:plas}b) and compare them to the calculated values of $\omega_p$ using Equation~(\ref{eq:plasmon}) and our device geometry parameters (red dashed lines). In this calculation, the effective width $w_e$ of the electron system and the corresponding central microchannel electron density $n_s$ are calculated using FEM for each value of $V_{\mathrm{ch}}$. As shown in Fig.~\ref{fig:plas}b, we find good agreement between our data and the two-dimensional screened plasmon model for the fundamental mode and its first six harmonics. The results reveal plasmon modes in a frequency range compatible with cQED systems, and that their frequency can be electrostatically tuned over an extremely broad range ($\simeq~2~-~3$~GHz) by controlling the areal density of electrons. We note that the most pronounced modes in Fig.~\ref{fig:plas}a correspond to odd harmonics. This effect can be understood as arising from a preferential coupling between the microwave excitation field and plasmon modes that have charge density nodes at the ends of the microchannel (see Supplementary Information, section III).

The data show the resonances appearing only in regime II, and not in the low-conductivity Wigner solid (regime III). This is consistent with a significant reduction of the charge density oscillation frequency arising from the phononic modes of the crystal when they are coupled to the elementary excitations of the helium surface (ripplons). In the long wavelength limit, this coupling to ripplons reduces the bare longitudinal plasmon frequency by a factor of $\sqrt{m/m^{*}}$~\cite{fisher1979phonon}, where $m^{*} \gtrsim 100 m_e$ parameterizes the effective mass of electrons on helium in the Wigner solid state~\cite{StanDahm1989}. Finally, we note the data in Fig.~\ref{fig:plas}a show the transition to the low-conductivity Wigner solid state is significantly non-uniform as a function of microwave drive frequency. This non-resonant effect could be associated with the absorption of microwave energy by other parts of the device, e.g. the resist layer between the top and bottom electrodes or variable microwave transmission due to impedance mismatches in the drive line.

\vspace{0.5cm}

\noindent \textbf{Power dependence}

\begin{figure}[h]
    \centering
    \includegraphics[width=0.8\textwidth]{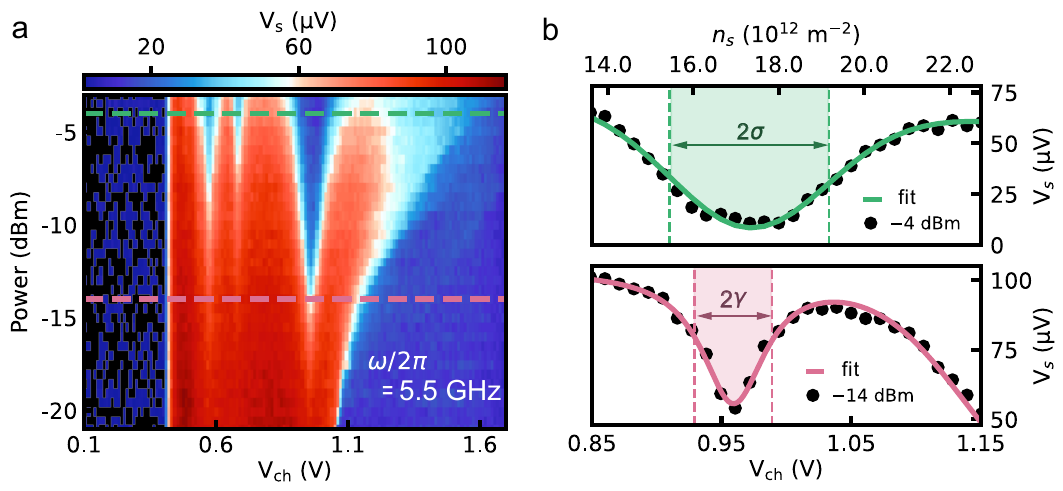} 
    \caption{\textbf{Plasmon power dependence}. (a) Microwave power dependent transport measurements at fixed excitation frequency {$\omega/2\pi~=~5.5$~GHz}. With increasing power the resonant plasmon modes broaden significantly. Measurements were performed at {$V_{\mathrm{ac}}~=~6$~mV}, {$f_{\mathrm{ac}}~=~3$~MHz}, {$V_{\mathrm{gt}}~=~0$~V}, {$V_{\mathrm{res}}~=~0.4$~V}, and {$T=18$~mK}. (b) Linecuts of the $n=3$ plasmon mode, corresponding to the horizontal dashed lines in (a). Depending on the applied microwave power, the plasmon resonances are fit to either a Lorentzian or Gaussian, superimposed on a smoothly varying background function (see Supplementary Information, section IV).}
    \label{fig:power}
\end{figure}

\noindent The measurements presented in Fig.~\ref{fig:power}a show how increasing microwave power modifies the transport characteristics and plasmon response at a fixed microwave excitation frequency of 5.5 GHz. Here, the transition between the high-conductivity (regime II) and Wigner solid (regime III) states shifts to higher electron density as the microwave power increases, indicating a weakening of the positional order of the Wigner solid, as discussed earlier. Additionally, as the channel density is varied, the $5.5$~GHz microwave drive generates the $n = 7, 5$, and $3$ plasmon modes (from left to right, respectively), which manifest and broaden with increasing power. As it is the most prominent, we will focus on the $n = 3$ mode centered at $V_{\mathrm{ch}}~\simeq~0.96$~V for the remainder of the analysis in this section. In Fig.~\ref{fig:power}b, we show linecuts of the data at low and high microwave power. At low power ($P = -14$~dBm), we find the data is well described by a Lorentzian of width $2\gamma$ while at high power ($P = -4$~dBm) the resonance is captured by a Gaussian having a linewidth of $2\sigma$ (see Fig.~\ref{fig:power}b and Supplementary Information, section~IV). In general, the spectral linewidth and its power dependence contain information about intrinsic and inhomogeneous sources of plasmon broadening. In what follows, we discuss possible sources of broadening including static and dynamic density inhomogeneity of the charge carriers in and around the channel as well as plasmon energy loss.

To begin, the application of the microwave field can lead to a non-equilibrium redistribution of charge carriers inside the channel and could be responsible for the plasmon broadening we observe with increasing microwave power. In fact, previous experiments with electrons on helium have demonstrated a transient redistribution of electrons arising from resonant photovoltaic effects~\cite{Konstantinov2012}. We also note that the long and narrow aspect ratio of the central channel creates an electron density profile that varies transverse to the length of the channel as shown in Fig.~\ref{fig:dev1}b, where we plot the density along the y-direction determined from FEM. If we ascribe the broadening of the low power plasmon spectrum exclusively to static inhomogeneity in the electron density, the observed linewidth of the resonance feature corresponds to a density variation of $1.8 \times 10^{12}~\text{m}^{-2}$. This is significantly smaller than the full density variation transverse to the channel, indicating that a static density inhomogeneity is likely not the dominant mechanism contributing to the plasmon broadening at low powers. However, it is possible that non-uniform heating of the electron system by the microwave excitation could enhance static density inhomogeneities in the channel and contribute to the plasmon broadening we observe with increasing power.

Additionally, collective excitations such as plasmons are naturally sensitive to the boundary conditions imposed by their environment. The boundary conditions for longitudinal plasmons excited along the channel are determined by the difference in density and conductivity between the electrons in the central channel and those in the reservoir regions of the device. At the ends of the central channel, the electron density varies smoothly over a distance of approximately $\delta L = 10~\upmu$m, leading to an uncertainty in the plasmon wavelength. Using Equation~(\ref{eq:plasmon}) the corresponding uncertainty in the plasmon frequency would be approximately 400~MHz, which is the same order of magnitude as the plasmon linewidth we observe at low microwave power. Furthermore, as the microwave drive power is increased, additional broadening could result from a dynamical redistribution of charge carriers in the vicinity of this boundary. Another possible broadening mechanism is the finite transparency at the boundary on either end of the microchannel. In this scenario, plasmon damping results from a leakage of the charge density wave from the central channel to the electron system in the reservoirs and subsequent thermalization~\cite{Kapralov2020}. In addition, it has been proposed that plasmon damping can arise from the difference in the conductivity between the channel and reservoir electron systems~\cite{Satou2005}.

Lastly, it is also important to consider the intrinsic energy losses of the plasmon modes. In this system, energy loss can arise from the screening currents in the lossy resistive metallic gate electrodes, as well as from interactions with the helium surface vibrational modes (ripplons) and phonons in the liquid, which are the dominant energy and momentum relaxation mechanisms at the relevant experimental temperatures ($T~\lesssim~0.8$~K)~\cite{dykman2003, rees2020dynamical}. In the absence of inhomogeneous effects these losses should dominate the plasmon linewidth. If we assume that the low power ($P = -14$~dBm) broadening arises predominately from intrinsic losses, we estimate a lower bound for the plasmon lifetime $\tau_p \sim 1$~ns.

In summary, we have demonstrated a microchannel device architecture that enables us to precisely engineer spatially-confined microwave frequency plasmonic modes in electrons on helium. The generation of these plasmons resonantly drives the electron system out of equilibrium, which we detect via low-frequency ac transport measurements of the device conductivity. Precise control over the electrostatic environment of the microchannel-confined electrons enables a tunability of the plasmon modes over a frequency range of $\sim 3$~GHz and we find good agreement between the observed plasmonic mode structure and the two-dimensional screened plasmon dispersion relation. Power-dependent measurements allow us to explore the interplay between the microwave drive and the non-equilibrium plasmon response in the device and could open the door to investigate plasmon dynamics in the hydrodynamic regime~\cite{Torre2019}.

The high degree of spatial control and broad microwave frequency tunability provided by this type of microchannel device offers a compelling framework for integrating charged collective oscillations of electrons on helium with circuit quantum electrodynamic systems. Devices utilizing an improved and optimized microwave environment will ultimately be needed for future cQED experiments with ensembles of electrons on helium. Placing many-electron on helium systems into high-quality factor microwave cavities opens entirely new avenues for exploring cavity optoplasmonics with collective modes in both the Coulomb liquid and solid phases. Similarly, integration with charge sensitive superconducting qubits~\cite{serniak2019direct} would enable fast readout of individual and collective electron dynamics and could be used to reveal the microscopic breakdown in the coupling of the electrons to the quantum field of helium surface waves~\cite{rees2016stick, Dykman2016Phys}. Alternatively, hybrid systems composed of electrons on helium coupled to superconducting qubits could be used as a model system for understanding qubit decoherence produced by charged fluctuators~\cite{martinis2005decoherence, klimov2018fluctuations} in a systematic and tunable fashion. 

\vspace{0.5cm}

\noindent \textbf{\large Methods}

\noindent \textbf{Microchannel device fabrication}

\noindent The microchannel device was fabricated on high resistivity silicon, using a combination of both photolithography and electron beam lithography. The first layer of bottom electrodes was fabricated using an image reversal resist AZ5214E, which was spun on and pre-baked ($90^{\circ}$~C for 30~min), followed by a short UV-light exposure (3~s) to define the electrode pattern. The chip is then post-baked (30~min at $95^{\circ}$~C), followed by a flood exposure (24~s), and developed for 45~s in AZ300MIF, and finally rinsed for 15~s in deionized (DI) water. A combination of 3~nm of Ti followed by 40~nm of Au is then deposited via thermal evaporation, followed by lift-off. The central channel electrode is patterned via electron beam lithography using PMMA-C2 resist and 4~nm-50~nm Ti-Au bilayer deposition. Next we form the dielectric layer with S1813 photoresist, hard-baked at $200^{\circ}$~C. The final helium channel depth is defined by the thickness of this dielectric layer. The 1.4~$\upmu$m deep channels are realized by spin coating resist three times at 4000~RPM for 60~s. The top layer guard and gate electrodes are then patternedsimilarly to the bottom electrodes. Finally, to remove all excess resist from the channels, the sample is oxygen plasma etched.

\vspace{0.3cm}

\noindent \textbf{Microchannel device assembly and packaging}

\noindent The microchannel device is wire-bonded onto a custom-made printed circuit board (PCB) that is placed into a superfluid-leak-tight copper sample box that mounts to the mixing chamber plate of a cryogen-free dilution refrigerator. A stainless steel capillary line is hard soldered into the sample box, which extends to a room temperature volume containing helium gas for filling the microchannels with liquid. The sample box contains a tungsten filament for thermionic electron emission, which is achieved by applying a $-2$~V amplitude, $300$~ms duration, square pulse to the filament.

\vspace{0.3cm}

\noindent \textbf{Transport measurements}

\noindent The transport measurements discussed here are performed as follows. The left and right reservoir electrodes are both biased with the same positive dc voltage ($V_{\mathrm{res}}$) while an ac drive voltage ($V_{\mathrm{ac}}$) in the {1~MHz~$-$~3~MHz} frequency range ($f_{\mathrm{ac}}$) is simultaneously applied to the left side reservoir electrode to drive electrons through the central channel. The electron transport is characterized by sweeping the channel electrode voltage $V_{\mathrm{ch}}$ while simultaneously measuring the transport signal ($V_\textrm{s}$) induced on the right side reservoir via phase-sensitive lock-in techniques.

\bibliography{pme_bib}

\begin{thebibliography}{73}%
\makeatletter
\providecommand \@ifxundefined [1]{%
 \@ifx{#1\undefined}
}%
\providecommand \@ifnum [1]{%
 \ifnum #1\expandafter \@firstoftwo
 \else \expandafter \@secondoftwo
 \fi
}%
\providecommand \@ifx [1]{%
 \ifx #1\expandafter \@firstoftwo
 \else \expandafter \@secondoftwo
 \fi
}%
\providecommand \natexlab [1]{#1}%
\providecommand \enquote  [1]{``#1''}%
\providecommand \bibnamefont  [1]{#1}%
\providecommand \bibfnamefont [1]{#1}%
\providecommand \citenamefont [1]{#1}%
\providecommand \href@noop [0]{\@secondoftwo}%
\providecommand \href [0]{\begingroup \@sanitize@url \@href}%
\providecommand \@href[1]{\@@startlink{#1}\@@href}%
\providecommand \@@href[1]{\endgroup#1\@@endlink}%
\providecommand \@sanitize@url [0]{\catcode `\\12\catcode `\$12\catcode `\&12\catcode `\#12\catcode `\^12\catcode `\_12\catcode `\%12\relax}%
\providecommand \@@startlink[1]{}%
\providecommand \@@endlink[0]{}%
\providecommand \url  [0]{\begingroup\@sanitize@url \@url }%
\providecommand \@url [1]{\endgroup\@href {#1}{\urlprefix }}%
\providecommand \urlprefix  [0]{URL }%
\providecommand \Eprint [0]{\href }%
\providecommand \doibase [0]{https://doi.org/}%
\providecommand \selectlanguage [0]{\@gobble}%
\providecommand \bibinfo  [0]{\@secondoftwo}%
\providecommand \bibfield  [0]{\@secondoftwo}%
\providecommand \translation [1]{[#1]}%
\providecommand \BibitemOpen [0]{}%
\providecommand \bibitemStop [0]{}%
\providecommand \bibitemNoStop [0]{.\EOS\space}%
\providecommand \EOS [0]{\spacefactor3000\relax}%
\providecommand \BibitemShut  [1]{\csname bibitem#1\endcsname}%
\let\auto@bib@innerbib\@empty
\bibitem [{\citenamefont {Blais}\ \emph {et~al.}(2021)\citenamefont {Blais}, \citenamefont {Grimsmo}, \citenamefont {Girvin},\ and\ \citenamefont {Wallraff}}]{Blais2021}%
  \BibitemOpen
  \bibfield  {author} {\bibinfo {author} {\bibfnamefont {A.}~\bibnamefont {Blais}}, \bibinfo {author} {\bibfnamefont {A.~L.}\ \bibnamefont {Grimsmo}}, \bibinfo {author} {\bibfnamefont {S.}~\bibnamefont {Girvin}},\ and\ \bibinfo {author} {\bibfnamefont {A.}~\bibnamefont {Wallraff}},\ }\bibfield  {title} {\bibinfo {title} {Circuit quantum electrodynamics},\ }\href {https://doi.org/10.1103/RevModPhys.93.025005} {\bibfield  {journal} {\bibinfo  {journal} {Rev. Mod. Phys.}\ }\textbf {\bibinfo {volume} {93}},\ \bibinfo {pages} {025005} (\bibinfo {year} {2021})}\BibitemShut {NoStop}%
\bibitem [{\citenamefont {Blais}\ \emph {et~al.}(2004)\citenamefont {Blais}, \citenamefont {Huang}, \citenamefont {Wallraff}, \citenamefont {Girvin},\ and\ \citenamefont {Schoelkopf}}]{Blais2004}%
  \BibitemOpen
  \bibfield  {author} {\bibinfo {author} {\bibfnamefont {A.}~\bibnamefont {Blais}}, \bibinfo {author} {\bibfnamefont {R.-S.}\ \bibnamefont {Huang}}, \bibinfo {author} {\bibfnamefont {A.}~\bibnamefont {Wallraff}}, \bibinfo {author} {\bibfnamefont {S.~M.}\ \bibnamefont {Girvin}},\ and\ \bibinfo {author} {\bibfnamefont {R.~J.}\ \bibnamefont {Schoelkopf}},\ }\bibfield  {title} {\bibinfo {title} {Cavity quantum electrodynamics for superconducting electrical circuits: An architecture for quantum computation},\ }\href {https://doi.org/10.1103/PhysRevA.69.062320} {\bibfield  {journal} {\bibinfo  {journal} {Phys. Rev. A}\ }\textbf {\bibinfo {volume} {69}},\ \bibinfo {pages} {062320} (\bibinfo {year} {2004})}\BibitemShut {NoStop}%
\bibitem [{\citenamefont {Petersson}\ \emph {et~al.}(2012)\citenamefont {Petersson}, \citenamefont {McFaul}, \citenamefont {Schroer}, \citenamefont {Jung}, \citenamefont {Taylor}, \citenamefont {Houck},\ and\ \citenamefont {Petta}}]{Petersson2012}%
  \BibitemOpen
  \bibfield  {author} {\bibinfo {author} {\bibfnamefont {K.}~\bibnamefont {Petersson}}, \bibinfo {author} {\bibfnamefont {L.}~\bibnamefont {McFaul}}, \bibinfo {author} {\bibfnamefont {M.}~\bibnamefont {Schroer}}, \bibinfo {author} {\bibfnamefont {M.}~\bibnamefont {Jung}}, \bibinfo {author} {\bibfnamefont {J.}~\bibnamefont {Taylor}}, \bibinfo {author} {\bibfnamefont {A.}~\bibnamefont {Houck}},\ and\ \bibinfo {author} {\bibfnamefont {J.}~\bibnamefont {Petta}},\ }\bibfield  {title} {\bibinfo {title} {Circuit quantum electrodynamics with a spin qubit},\ }\href {https://doi.org/10.1038/nature11559} {\bibfield  {journal} {\bibinfo  {journal} {Nature}\ }\textbf {\bibinfo {volume} {490}},\ \bibinfo {pages} {380} (\bibinfo {year} {2012})}\BibitemShut {NoStop}%
\bibitem [{\citenamefont {Koolstra}\ \emph {et~al.}(2019)\citenamefont {Koolstra}, \citenamefont {Yang},\ and\ \citenamefont {Schuster}}]{koolstra2019coupling}%
  \BibitemOpen
  \bibfield  {author} {\bibinfo {author} {\bibfnamefont {G.}~\bibnamefont {Koolstra}}, \bibinfo {author} {\bibfnamefont {G.}~\bibnamefont {Yang}},\ and\ \bibinfo {author} {\bibfnamefont {D.}~\bibnamefont {Schuster}},\ }\bibfield  {title} {\bibinfo {title} {Coupling a single electron on superfluid helium to a superconducting resonator},\ }\href {https://doi.org/10.1038/s41467-019-13335-7} {\bibfield  {journal} {\bibinfo  {journal} {Nat. Commun}\ }\textbf {\bibinfo {volume} {10}},\ \bibinfo {pages} {5323} (\bibinfo {year} {2019})}\BibitemShut {NoStop}%
\bibitem [{\citenamefont {Zhou}\ \emph {et~al.}(2022)\citenamefont {Zhou}, \citenamefont {Koolstra}, \citenamefont {Zhang}, \citenamefont {Yang}, \citenamefont {Han}, \citenamefont {Dizdar}, \citenamefont {Li}, \citenamefont {Divan}, \citenamefont {Guo}, \citenamefont {Murch}, \citenamefont {Schuster},\ and\ \citenamefont {Jin}}]{zhou2022single}%
  \BibitemOpen
  \bibfield  {author} {\bibinfo {author} {\bibfnamefont {X.}~\bibnamefont {Zhou}}, \bibinfo {author} {\bibfnamefont {G.}~\bibnamefont {Koolstra}}, \bibinfo {author} {\bibfnamefont {X.}~\bibnamefont {Zhang}}, \bibinfo {author} {\bibfnamefont {G.}~\bibnamefont {Yang}}, \bibinfo {author} {\bibfnamefont {X.}~\bibnamefont {Han}}, \bibinfo {author} {\bibfnamefont {B.}~\bibnamefont {Dizdar}}, \bibinfo {author} {\bibfnamefont {X.}~\bibnamefont {Li}}, \bibinfo {author} {\bibfnamefont {R.}~\bibnamefont {Divan}}, \bibinfo {author} {\bibfnamefont {W.}~\bibnamefont {Guo}}, \bibinfo {author} {\bibfnamefont {K.~W.}\ \bibnamefont {Murch}}, \bibinfo {author} {\bibfnamefont {D.~I.}\ \bibnamefont {Schuster}},\ and\ \bibinfo {author} {\bibfnamefont {D.}~\bibnamefont {Jin}},\ }\bibfield  {title} {\bibinfo {title} {Single electrons on solid neon as a solid-state qubit platform},\ }\href {https://doi.org/10.1038/s41586-022-04539-x} {\bibfield  {journal} {\bibinfo  {journal} {Nature}\ }\textbf {\bibinfo {volume} {605}},\ \bibinfo
  {pages} {46} (\bibinfo {year} {2022})}\BibitemShut {NoStop}%
\bibitem [{\citenamefont {Zhou}\ \emph {et~al.}(2024)\citenamefont {Zhou}, \citenamefont {Li}, \citenamefont {Chen}, \citenamefont {Koolstra}, \citenamefont {Yang}, \citenamefont {Dizdar}, \citenamefont {Huang}, \citenamefont {Wang}, \citenamefont {Han}, \citenamefont {Zhang}, \citenamefont {Schuster},\ and\ \citenamefont {Jin}}]{Zhou2023}%
  \BibitemOpen
  \bibfield  {author} {\bibinfo {author} {\bibfnamefont {X.}~\bibnamefont {Zhou}}, \bibinfo {author} {\bibfnamefont {X.}~\bibnamefont {Li}}, \bibinfo {author} {\bibfnamefont {Q.}~\bibnamefont {Chen}}, \bibinfo {author} {\bibfnamefont {G.}~\bibnamefont {Koolstra}}, \bibinfo {author} {\bibfnamefont {G.}~\bibnamefont {Yang}}, \bibinfo {author} {\bibfnamefont {B.}~\bibnamefont {Dizdar}}, \bibinfo {author} {\bibfnamefont {Y.}~\bibnamefont {Huang}}, \bibinfo {author} {\bibfnamefont {C.~S.}\ \bibnamefont {Wang}}, \bibinfo {author} {\bibfnamefont {X.}~\bibnamefont {Han}}, \bibinfo {author} {\bibfnamefont {X.}~\bibnamefont {Zhang}}, \bibinfo {author} {\bibfnamefont {D.~I.}\ \bibnamefont {Schuster}},\ and\ \bibinfo {author} {\bibfnamefont {D.}~\bibnamefont {Jin}},\ }\bibfield  {title} {\bibinfo {title} {Electron charge qubit with 0.1 millisecond coherence time},\ }\href {https://doi.org/10.1038/s41567-023-02247-5} {\bibfield  {journal} {\bibinfo  {journal} {Nat. Phys}\ }\textbf {\bibinfo {volume} {20}},\ \bibinfo
  {pages} {116} (\bibinfo {year} {2024})}\BibitemShut {NoStop}%
\bibitem [{\citenamefont {LaHaye}\ \emph {et~al.}(2009)\citenamefont {LaHaye}, \citenamefont {Suh}, \citenamefont {Echternach}, \citenamefont {Schwab},\ and\ \citenamefont {Roukes}}]{lahaye2009nanomechanical}%
  \BibitemOpen
  \bibfield  {author} {\bibinfo {author} {\bibfnamefont {M.}~\bibnamefont {LaHaye}}, \bibinfo {author} {\bibfnamefont {J.}~\bibnamefont {Suh}}, \bibinfo {author} {\bibfnamefont {P.}~\bibnamefont {Echternach}}, \bibinfo {author} {\bibfnamefont {K.~C.}\ \bibnamefont {Schwab}},\ and\ \bibinfo {author} {\bibfnamefont {M.~L.}\ \bibnamefont {Roukes}},\ }\bibfield  {title} {\bibinfo {title} {Nanomechanical measurements of a superconducting qubit},\ }\href {https://doi.org/10.1038/nature08093} {\bibfield  {journal} {\bibinfo  {journal} {Nature}\ }\textbf {\bibinfo {volume} {459}},\ \bibinfo {pages} {960} (\bibinfo {year} {2009})}\BibitemShut {NoStop}%
\bibitem [{\citenamefont {O`Connell}\ \emph {et~al.}(2010)\citenamefont {O`Connell}, \citenamefont {Hofheinz}, \citenamefont {Ansmann}, \citenamefont {Bialczak}, \citenamefont {Lenander}, \citenamefont {Lucero}, \citenamefont {Neeley}, \citenamefont {Sank}, \citenamefont {Wang}, \citenamefont {Weides}, \citenamefont {Wenner}, \citenamefont {Martinis},\ and\ \citenamefont {Cleland}}]{o2010quantum}%
  \BibitemOpen
  \bibfield  {author} {\bibinfo {author} {\bibfnamefont {A.~D.}\ \bibnamefont {O`Connell}}, \bibinfo {author} {\bibfnamefont {M.}~\bibnamefont {Hofheinz}}, \bibinfo {author} {\bibfnamefont {M.}~\bibnamefont {Ansmann}}, \bibinfo {author} {\bibfnamefont {R.~C.}\ \bibnamefont {Bialczak}}, \bibinfo {author} {\bibfnamefont {M.}~\bibnamefont {Lenander}}, \bibinfo {author} {\bibfnamefont {E.}~\bibnamefont {Lucero}}, \bibinfo {author} {\bibfnamefont {M.}~\bibnamefont {Neeley}}, \bibinfo {author} {\bibfnamefont {D.}~\bibnamefont {Sank}}, \bibinfo {author} {\bibfnamefont {H.}~\bibnamefont {Wang}}, \bibinfo {author} {\bibfnamefont {M.}~\bibnamefont {Weides}}, \bibinfo {author} {\bibfnamefont {J.}~\bibnamefont {Wenner}}, \bibinfo {author} {\bibfnamefont {J.~M.}\ \bibnamefont {Martinis}},\ and\ \bibinfo {author} {\bibfnamefont {A.}~\bibnamefont {Cleland}},\ }\bibfield  {title} {\bibinfo {title} {Quantum ground state and single-phonon control of a mechanical resonator},\ }\href {https://doi.org/10.1038/nature08967}
  {\bibfield  {journal} {\bibinfo  {journal} {Nature}\ }\textbf {\bibinfo {volume} {464}},\ \bibinfo {pages} {697} (\bibinfo {year} {2010})}\BibitemShut {NoStop}%
\bibitem [{\citenamefont {Pirkkalainen}\ \emph {et~al.}(2013)\citenamefont {Pirkkalainen}, \citenamefont {Cho}, \citenamefont {Li}, \citenamefont {Paraoanu}, \citenamefont {Hakonen},\ and\ \citenamefont {Sillanp{\"a}{\"a}}}]{Pirkkalainen2013}%
  \BibitemOpen
  \bibfield  {author} {\bibinfo {author} {\bibfnamefont {J.}~\bibnamefont {Pirkkalainen}}, \bibinfo {author} {\bibfnamefont {S.}~\bibnamefont {Cho}}, \bibinfo {author} {\bibfnamefont {J.}~\bibnamefont {Li}}, \bibinfo {author} {\bibfnamefont {G.}~\bibnamefont {Paraoanu}}, \bibinfo {author} {\bibfnamefont {P.}~\bibnamefont {Hakonen}},\ and\ \bibinfo {author} {\bibfnamefont {M.}~\bibnamefont {Sillanp{\"a}{\"a}}},\ }\bibfield  {title} {\bibinfo {title} {Hybrid circuit cavity quantum electrodynamics with a micromechanical resonator},\ }\href {https://doi.org/10.1038/nature11821} {\bibfield  {journal} {\bibinfo  {journal} {Nature}\ }\textbf {\bibinfo {volume} {494}},\ \bibinfo {pages} {211} (\bibinfo {year} {2013})}\BibitemShut {NoStop}%
\bibitem [{\citenamefont {Clerk}\ \emph {et~al.}(2020)\citenamefont {Clerk}, \citenamefont {Lehnert}, \citenamefont {Bertet}, \citenamefont {Petta},\ and\ \citenamefont {Nakamura}}]{clerk2020hybrid}%
  \BibitemOpen
  \bibfield  {author} {\bibinfo {author} {\bibfnamefont {A.}~\bibnamefont {Clerk}}, \bibinfo {author} {\bibfnamefont {K.}~\bibnamefont {Lehnert}}, \bibinfo {author} {\bibfnamefont {P.}~\bibnamefont {Bertet}}, \bibinfo {author} {\bibfnamefont {J.}~\bibnamefont {Petta}},\ and\ \bibinfo {author} {\bibfnamefont {Y.}~\bibnamefont {Nakamura}},\ }\bibfield  {title} {\bibinfo {title} {Hybrid quantum systems with circuit quantum electrodynamics},\ }\href {https://doi.org/10.1038/s41567-020-0797-9} {\bibfield  {journal} {\bibinfo  {journal} {Nat. Phys}\ }\textbf {\bibinfo {volume} {16}},\ \bibinfo {pages} {257} (\bibinfo {year} {2020})}\BibitemShut {NoStop}%
\bibitem [{\citenamefont {Goryachev}\ \emph {et~al.}(2014)\citenamefont {Goryachev}, \citenamefont {Farr}, \citenamefont {Creedon}, \citenamefont {Fan}, \citenamefont {Kostylev},\ and\ \citenamefont {Tobar}}]{goryachev2014high}%
  \BibitemOpen
  \bibfield  {author} {\bibinfo {author} {\bibfnamefont {M.}~\bibnamefont {Goryachev}}, \bibinfo {author} {\bibfnamefont {W.~G.}\ \bibnamefont {Farr}}, \bibinfo {author} {\bibfnamefont {D.~L.}\ \bibnamefont {Creedon}}, \bibinfo {author} {\bibfnamefont {Y.}~\bibnamefont {Fan}}, \bibinfo {author} {\bibfnamefont {M.}~\bibnamefont {Kostylev}},\ and\ \bibinfo {author} {\bibfnamefont {M.~E.}\ \bibnamefont {Tobar}},\ }\bibfield  {title} {\bibinfo {title} {High-cooperativity cavity {QED} with magnons at microwave frequencies},\ }\href {https://doi.org/10.1103/PhysRevApplied.2.054002} {\bibfield  {journal} {\bibinfo  {journal} {Phys. Rev. Appl}\ }\textbf {\bibinfo {volume} {2}},\ \bibinfo {pages} {054002} (\bibinfo {year} {2014})}\BibitemShut {NoStop}%
\bibitem [{\citenamefont {Tabuchi}\ \emph {et~al.}(2015)\citenamefont {Tabuchi}, \citenamefont {Ishino}, \citenamefont {Noguchi}, \citenamefont {Ishikawa}, \citenamefont {Yamazaki}, \citenamefont {Usami},\ and\ \citenamefont {Nakamura}}]{tabuchi2015coherent}%
  \BibitemOpen
  \bibfield  {author} {\bibinfo {author} {\bibfnamefont {Y.}~\bibnamefont {Tabuchi}}, \bibinfo {author} {\bibfnamefont {S.}~\bibnamefont {Ishino}}, \bibinfo {author} {\bibfnamefont {A.}~\bibnamefont {Noguchi}}, \bibinfo {author} {\bibfnamefont {T.}~\bibnamefont {Ishikawa}}, \bibinfo {author} {\bibfnamefont {R.}~\bibnamefont {Yamazaki}}, \bibinfo {author} {\bibfnamefont {K.}~\bibnamefont {Usami}},\ and\ \bibinfo {author} {\bibfnamefont {Y.}~\bibnamefont {Nakamura}},\ }\bibfield  {title} {\bibinfo {title} {Coherent coupling between a ferromagnetic magnon and a superconducting qubit},\ }\href {https://doi.org/10.1126/science.aaa3693} {\bibfield  {journal} {\bibinfo  {journal} {Science}\ }\textbf {\bibinfo {volume} {349}},\ \bibinfo {pages} {405} (\bibinfo {year} {2015})}\BibitemShut {NoStop}%
\bibitem [{\citenamefont {Li}\ \emph {et~al.}(2018)\citenamefont {Li}, \citenamefont {Zhu},\ and\ \citenamefont {Agarwal}}]{li2018magnon}%
  \BibitemOpen
  \bibfield  {author} {\bibinfo {author} {\bibfnamefont {J.}~\bibnamefont {Li}}, \bibinfo {author} {\bibfnamefont {S.-Y.}\ \bibnamefont {Zhu}},\ and\ \bibinfo {author} {\bibfnamefont {G.}~\bibnamefont {Agarwal}},\ }\bibfield  {title} {\bibinfo {title} {Magnon-photon-phonon entanglement in cavity magnomechanics},\ }\href {https://doi.org/10.1103/PhysRevLett.121.203601} {\bibfield  {journal} {\bibinfo  {journal} {Phys. Rev. Lett}\ }\textbf {\bibinfo {volume} {121}},\ \bibinfo {pages} {203601} (\bibinfo {year} {2018})}\BibitemShut {NoStop}%
\bibitem [{\citenamefont {Lachance-Quirion}\ \emph {et~al.}(2020)\citenamefont {Lachance-Quirion}, \citenamefont {Wolski}, \citenamefont {Tabuchi}, \citenamefont {Kono}, \citenamefont {Usami},\ and\ \citenamefont {Nakamura}}]{lachance2020entanglement}%
  \BibitemOpen
  \bibfield  {author} {\bibinfo {author} {\bibfnamefont {D.}~\bibnamefont {Lachance-Quirion}}, \bibinfo {author} {\bibfnamefont {S.~P.}\ \bibnamefont {Wolski}}, \bibinfo {author} {\bibfnamefont {Y.}~\bibnamefont {Tabuchi}}, \bibinfo {author} {\bibfnamefont {S.}~\bibnamefont {Kono}}, \bibinfo {author} {\bibfnamefont {K.}~\bibnamefont {Usami}},\ and\ \bibinfo {author} {\bibfnamefont {Y.}~\bibnamefont {Nakamura}},\ }\bibfield  {title} {\bibinfo {title} {Entanglement-based single-shot detection of a single magnon with a superconducting qubit},\ }\href {https://doi.org/10.1126/science.aaz9236} {\bibfield  {journal} {\bibinfo  {journal} {Science}\ }\textbf {\bibinfo {volume} {367}},\ \bibinfo {pages} {425} (\bibinfo {year} {2020})}\BibitemShut {NoStop}%
\bibitem [{\citenamefont {Gustafsson}\ \emph {et~al.}(2014)\citenamefont {Gustafsson}, \citenamefont {Aref}, \citenamefont {Kockum}, \citenamefont {Ekstr{\"o}m}, \citenamefont {Johansson},\ and\ \citenamefont {Delsing}}]{gustafsson2014propagating}%
  \BibitemOpen
  \bibfield  {author} {\bibinfo {author} {\bibfnamefont {M.~V.}\ \bibnamefont {Gustafsson}}, \bibinfo {author} {\bibfnamefont {T.}~\bibnamefont {Aref}}, \bibinfo {author} {\bibfnamefont {A.~F.}\ \bibnamefont {Kockum}}, \bibinfo {author} {\bibfnamefont {M.~K.}\ \bibnamefont {Ekstr{\"o}m}}, \bibinfo {author} {\bibfnamefont {G.}~\bibnamefont {Johansson}},\ and\ \bibinfo {author} {\bibfnamefont {P.}~\bibnamefont {Delsing}},\ }\bibfield  {title} {\bibinfo {title} {Propagating phonons coupled to an artificial atom},\ }\href {https://doi.org/10.1126/science.1257219} {\bibfield  {journal} {\bibinfo  {journal} {Science}\ }\textbf {\bibinfo {volume} {346}},\ \bibinfo {pages} {207} (\bibinfo {year} {2014})}\BibitemShut {NoStop}%
\bibitem [{\citenamefont {Chu}\ \emph {et~al.}(2017)\citenamefont {Chu}, \citenamefont {Kharel}, \citenamefont {Renninger}, \citenamefont {Burkhart}, \citenamefont {Frunzio}, \citenamefont {Rakich},\ and\ \citenamefont {Schoelkopf}}]{chu2017quantum}%
  \BibitemOpen
  \bibfield  {author} {\bibinfo {author} {\bibfnamefont {Y.}~\bibnamefont {Chu}}, \bibinfo {author} {\bibfnamefont {P.}~\bibnamefont {Kharel}}, \bibinfo {author} {\bibfnamefont {W.~H.}\ \bibnamefont {Renninger}}, \bibinfo {author} {\bibfnamefont {L.~D.}\ \bibnamefont {Burkhart}}, \bibinfo {author} {\bibfnamefont {L.}~\bibnamefont {Frunzio}}, \bibinfo {author} {\bibfnamefont {P.~T.}\ \bibnamefont {Rakich}},\ and\ \bibinfo {author} {\bibfnamefont {R.~J.}\ \bibnamefont {Schoelkopf}},\ }\bibfield  {title} {\bibinfo {title} {Quantum acoustics with superconducting qubits},\ }\href {https://doi.org/10.1126/science.aao1511} {\bibfield  {journal} {\bibinfo  {journal} {Science}\ }\textbf {\bibinfo {volume} {358}},\ \bibinfo {pages} {199} (\bibinfo {year} {2017})}\BibitemShut {NoStop}%
\bibitem [{\citenamefont {Satzinger}\ \emph {et~al.}(2018)\citenamefont {Satzinger}, \citenamefont {Zhong}, \citenamefont {Chang}, \citenamefont {Peairs}, \citenamefont {Bienfait}, \citenamefont {Chou}, \citenamefont {Cleland}, \citenamefont {Conner}, \citenamefont {Dumur}, \citenamefont {Grebel}, \citenamefont {Gutierrez}, \citenamefont {November}, \citenamefont {Povey}, \citenamefont {Whiteley}, \citenamefont {Awschalom}, \citenamefont {Schuster},\ and\ \citenamefont {Cleland}}]{satzinger2018quantum}%
  \BibitemOpen
  \bibfield  {author} {\bibinfo {author} {\bibfnamefont {K.}~\bibnamefont {Satzinger}}, \bibinfo {author} {\bibfnamefont {Y.}~\bibnamefont {Zhong}}, \bibinfo {author} {\bibfnamefont {H.}~\bibnamefont {Chang}}, \bibinfo {author} {\bibfnamefont {G.}~\bibnamefont {Peairs}}, \bibinfo {author} {\bibfnamefont {A.}~\bibnamefont {Bienfait}}, \bibinfo {author} {\bibfnamefont {M.}~\bibnamefont {Chou}}, \bibinfo {author} {\bibfnamefont {A.}~\bibnamefont {Cleland}}, \bibinfo {author} {\bibfnamefont {C.}~\bibnamefont {Conner}}, \bibinfo {author} {\bibfnamefont {{\`E}.}~\bibnamefont {Dumur}}, \bibinfo {author} {\bibfnamefont {J.}~\bibnamefont {Grebel}}, \bibinfo {author} {\bibfnamefont {I.}~\bibnamefont {Gutierrez}}, \bibinfo {author} {\bibfnamefont {B.}~\bibnamefont {November}}, \bibinfo {author} {\bibfnamefont {R.}~\bibnamefont {Povey}}, \bibinfo {author} {\bibfnamefont {S.}~\bibnamefont {Whiteley}}, \bibinfo {author} {\bibfnamefont {D.}~\bibnamefont {Awschalom}}, \bibinfo {author} {\bibfnamefont {D.}~\bibnamefont
  {Schuster}},\ and\ \bibinfo {author} {\bibfnamefont {A.}~\bibnamefont {Cleland}},\ }\bibfield  {title} {\bibinfo {title} {Quantum control of surface acoustic-wave phonons},\ }\href {https://doi.org/10.1038/s41586-018-0719-5} {\bibfield  {journal} {\bibinfo  {journal} {Nature}\ }\textbf {\bibinfo {volume} {563}},\ \bibinfo {pages} {661} (\bibinfo {year} {2018})}\BibitemShut {NoStop}%
\bibitem [{\citenamefont {Kitzman}\ \emph {et~al.}(2023)\citenamefont {Kitzman}, \citenamefont {Lane}, \citenamefont {Undershute}, \citenamefont {Harrington}, \citenamefont {Beysengulov}, \citenamefont {Mikolas}, \citenamefont {Murch},\ and\ \citenamefont {Pollanen}}]{kitzman2023phononic}%
  \BibitemOpen
  \bibfield  {author} {\bibinfo {author} {\bibfnamefont {J.}~\bibnamefont {Kitzman}}, \bibinfo {author} {\bibfnamefont {J.}~\bibnamefont {Lane}}, \bibinfo {author} {\bibfnamefont {C.}~\bibnamefont {Undershute}}, \bibinfo {author} {\bibfnamefont {P.}~\bibnamefont {Harrington}}, \bibinfo {author} {\bibfnamefont {N.}~\bibnamefont {Beysengulov}}, \bibinfo {author} {\bibfnamefont {C.}~\bibnamefont {Mikolas}}, \bibinfo {author} {\bibfnamefont {K.}~\bibnamefont {Murch}},\ and\ \bibinfo {author} {\bibfnamefont {J.}~\bibnamefont {Pollanen}},\ }\bibfield  {title} {\bibinfo {title} {Phononic bath engineering of a superconducting qubit},\ }\href {https://doi.org/10.1038/s41467-023-39682-0} {\bibfield  {journal} {\bibinfo  {journal} {Nat. Commun}\ }\textbf {\bibinfo {volume} {14}},\ \bibinfo {pages} {3910} (\bibinfo {year} {2023})}\BibitemShut {NoStop}%
\bibitem [{\citenamefont {Kubo}\ \emph {et~al.}(2010)\citenamefont {Kubo}, \citenamefont {Ong}, \citenamefont {Bertet}, \citenamefont {Vion}, \citenamefont {Jacques}, \citenamefont {Zheng}, \citenamefont {Dr{\'e}au}, \citenamefont {Roch}, \citenamefont {Auff{\`e}ves}, \citenamefont {Jelezko}, \citenamefont {Wrachtrup}, \citenamefont {Barthe}, \citenamefont {Bergonzo},\ and\ \citenamefont {Esteve}}]{kubo2010strong}%
  \BibitemOpen
  \bibfield  {author} {\bibinfo {author} {\bibfnamefont {Y.}~\bibnamefont {Kubo}}, \bibinfo {author} {\bibfnamefont {F.~R.}\ \bibnamefont {Ong}}, \bibinfo {author} {\bibfnamefont {P.}~\bibnamefont {Bertet}}, \bibinfo {author} {\bibfnamefont {D.}~\bibnamefont {Vion}}, \bibinfo {author} {\bibfnamefont {V.}~\bibnamefont {Jacques}}, \bibinfo {author} {\bibfnamefont {D.}~\bibnamefont {Zheng}}, \bibinfo {author} {\bibfnamefont {A.}~\bibnamefont {Dr{\'e}au}}, \bibinfo {author} {\bibfnamefont {J.-F.}\ \bibnamefont {Roch}}, \bibinfo {author} {\bibfnamefont {A.}~\bibnamefont {Auff{\`e}ves}}, \bibinfo {author} {\bibfnamefont {F.}~\bibnamefont {Jelezko}}, \bibinfo {author} {\bibfnamefont {J.}~\bibnamefont {Wrachtrup}}, \bibinfo {author} {\bibfnamefont {M.~F.}\ \bibnamefont {Barthe}}, \bibinfo {author} {\bibfnamefont {P.}~\bibnamefont {Bergonzo}},\ and\ \bibinfo {author} {\bibfnamefont {D.}~\bibnamefont {Esteve}},\ }\bibfield  {title} {\bibinfo {title} {Strong coupling of a spin ensemble to a superconducting resonator},\
  }\href {https://doi.org/10.1103/PhysRevLett.105.140502} {\bibfield  {journal} {\bibinfo  {journal} {Phys.\ Rev.\ Lett.}\ }\textbf {\bibinfo {volume} {105}},\ \bibinfo {pages} {140502} (\bibinfo {year} {2010})}\BibitemShut {NoStop}%
\bibitem [{\citenamefont {Schuster}\ \emph {et~al.}(2010{\natexlab{a}})\citenamefont {Schuster}, \citenamefont {Sears}, \citenamefont {Ginossar}, \citenamefont {DiCarlo}, \citenamefont {Frunzio}, \citenamefont {Morton}, \citenamefont {Wu}, \citenamefont {Briggs}, \citenamefont {Buckley}, \citenamefont {Awschalom},\ and\ \citenamefont {Schoelkopf}}]{schuster2010high}%
  \BibitemOpen
  \bibfield  {author} {\bibinfo {author} {\bibfnamefont {D.}~\bibnamefont {Schuster}}, \bibinfo {author} {\bibfnamefont {A.}~\bibnamefont {Sears}}, \bibinfo {author} {\bibfnamefont {E.}~\bibnamefont {Ginossar}}, \bibinfo {author} {\bibfnamefont {L.}~\bibnamefont {DiCarlo}}, \bibinfo {author} {\bibfnamefont {L.}~\bibnamefont {Frunzio}}, \bibinfo {author} {\bibfnamefont {J.}~\bibnamefont {Morton}}, \bibinfo {author} {\bibfnamefont {H.}~\bibnamefont {Wu}}, \bibinfo {author} {\bibfnamefont {G.}~\bibnamefont {Briggs}}, \bibinfo {author} {\bibfnamefont {B.}~\bibnamefont {Buckley}}, \bibinfo {author} {\bibfnamefont {D.}~\bibnamefont {Awschalom}},\ and\ \bibinfo {author} {\bibfnamefont {R.}~\bibnamefont {Schoelkopf}},\ }\bibfield  {title} {\bibinfo {title} {High-cooperativity coupling of electron-spin ensembles to superconducting cavities},\ }\href {https://doi.org/10.1103/PhysRevLett.105.140501} {\bibfield  {journal} {\bibinfo  {journal} {Phys. Rev. Lett.}\ }\textbf {\bibinfo {volume} {105}},\ \bibinfo {pages}
  {140501} (\bibinfo {year} {2010}{\natexlab{a}})}\BibitemShut {NoStop}%
\bibitem [{\citenamefont {Platzman}\ and\ \citenamefont {Dykman}(1999)}]{platzman1999quantum}%
  \BibitemOpen
  \bibfield  {author} {\bibinfo {author} {\bibfnamefont {P.}~\bibnamefont {Platzman}}\ and\ \bibinfo {author} {\bibfnamefont {M.}~\bibnamefont {Dykman}},\ }\bibfield  {title} {\bibinfo {title} {Quantum computing with electrons floating on liquid helium},\ }\href {https://doi.org/10.1126/science.284.5422.1967} {\bibfield  {journal} {\bibinfo  {journal} {Science}\ }\textbf {\bibinfo {volume} {284}},\ \bibinfo {pages} {1967} (\bibinfo {year} {1999})}\BibitemShut {NoStop}%
\bibitem [{\citenamefont {Dykman}\ \emph {et~al.}(2023)\citenamefont {Dykman}, \citenamefont {Asban}, \citenamefont {Chen}, \citenamefont {Jin},\ and\ \citenamefont {Lyon}}]{dykman2023spin}%
  \BibitemOpen
  \bibfield  {author} {\bibinfo {author} {\bibfnamefont {M.}~\bibnamefont {Dykman}}, \bibinfo {author} {\bibfnamefont {O.}~\bibnamefont {Asban}}, \bibinfo {author} {\bibfnamefont {Q.}~\bibnamefont {Chen}}, \bibinfo {author} {\bibfnamefont {D.}~\bibnamefont {Jin}},\ and\ \bibinfo {author} {\bibfnamefont {S.}~\bibnamefont {Lyon}},\ }\bibfield  {title} {\bibinfo {title} {Spin dynamics in quantum dots on liquid helium},\ }\href {https://doi.org/10.1103/PhysRevB.107.035437} {\bibfield  {journal} {\bibinfo  {journal} {Phys. Rev. B}\ }\textbf {\bibinfo {volume} {107}},\ \bibinfo {pages} {035437} (\bibinfo {year} {2023})}\BibitemShut {NoStop}%
\bibitem [{\citenamefont {Lyon}(2006)}]{lyon2006spin}%
  \BibitemOpen
  \bibfield  {author} {\bibinfo {author} {\bibfnamefont {S.}~\bibnamefont {Lyon}},\ }\bibfield  {title} {\bibinfo {title} {Spin-based quantum computing using electrons on liquid helium},\ }\href {https://doi.org/10.1103/PhysRevA.74.052338} {\bibfield  {journal} {\bibinfo  {journal} {Phys. Rev. A}\ }\textbf {\bibinfo {volume} {74}},\ \bibinfo {pages} {052338} (\bibinfo {year} {2006})}\BibitemShut {NoStop}%
\bibitem [{\citenamefont {Beysengulov}\ \emph {et~al.}(2024)\citenamefont {Beysengulov}, \citenamefont {Sch\o{}yen}, \citenamefont {Bilek}, \citenamefont {Flaten}, \citenamefont {Leinonen}, \citenamefont {Hjorth-Jensen}, \citenamefont {Pollanen}, \citenamefont {Kristiansen}, \citenamefont {Stewart}, \citenamefont {Weidman},\ and\ \citenamefont {Wilson}}]{Beysengulov2024}%
  \BibitemOpen
  \bibfield  {author} {\bibinfo {author} {\bibfnamefont {N.~R.}\ \bibnamefont {Beysengulov}}, \bibinfo {author} {\bibfnamefont {O.~S.}\ \bibnamefont {Sch\o{}yen}}, \bibinfo {author} {\bibfnamefont {S.~D.}\ \bibnamefont {Bilek}}, \bibinfo {author} {\bibfnamefont {J.~B.}\ \bibnamefont {Flaten}}, \bibinfo {author} {\bibfnamefont {O.}~\bibnamefont {Leinonen}}, \bibinfo {author} {\bibfnamefont {M.}~\bibnamefont {Hjorth-Jensen}}, \bibinfo {author} {\bibfnamefont {J.}~\bibnamefont {Pollanen}}, \bibinfo {author} {\bibfnamefont {H.~E.}\ \bibnamefont {Kristiansen}}, \bibinfo {author} {\bibfnamefont {Z.~J.}\ \bibnamefont {Stewart}}, \bibinfo {author} {\bibfnamefont {J.~D.}\ \bibnamefont {Weidman}},\ and\ \bibinfo {author} {\bibfnamefont {A.~K.}\ \bibnamefont {Wilson}},\ }\bibfield  {title} {\bibinfo {title} {Coulomb interaction-driven entanglement of electrons on helium},\ }\href {https://doi.org/10.1103/PRXQuantum.5.030324} {\bibfield  {journal} {\bibinfo  {journal} {PRX Quantum}\ }\textbf {\bibinfo {volume} {5}},\
  \bibinfo {pages} {030324} (\bibinfo {year} {2024})}\BibitemShut {NoStop}%
\bibitem [{\citenamefont {Schuster}\ \emph {et~al.}(2010{\natexlab{b}})\citenamefont {Schuster}, \citenamefont {Fragner}, \citenamefont {Dykman}, \citenamefont {Lyon},\ and\ \citenamefont {Schoelkopf}}]{schuster2010proposal}%
  \BibitemOpen
  \bibfield  {author} {\bibinfo {author} {\bibfnamefont {D.}~\bibnamefont {Schuster}}, \bibinfo {author} {\bibfnamefont {A.}~\bibnamefont {Fragner}}, \bibinfo {author} {\bibfnamefont {M.}~\bibnamefont {Dykman}}, \bibinfo {author} {\bibfnamefont {S.}~\bibnamefont {Lyon}},\ and\ \bibinfo {author} {\bibfnamefont {R.}~\bibnamefont {Schoelkopf}},\ }\bibfield  {title} {\bibinfo {title} {Proposal for manipulating and detecting spin and orbital states of trapped electrons on helium using cavity quantum electrodynamics},\ }\href {https://doi.org/10.1103/PhysRevLett.105.040503} {\bibfield  {journal} {\bibinfo  {journal} {Phys. Rev. Lett}\ }\textbf {\bibinfo {volume} {105}},\ \bibinfo {pages} {040503} (\bibinfo {year} {2010}{\natexlab{b}})}\BibitemShut {NoStop}%
\bibitem [{\citenamefont {Grimes}\ and\ \citenamefont {Adams}(1976)}]{grimesadams}%
  \BibitemOpen
  \bibfield  {author} {\bibinfo {author} {\bibfnamefont {C.}~\bibnamefont {Grimes}}\ and\ \bibinfo {author} {\bibfnamefont {G.}~\bibnamefont {Adams}},\ }\bibfield  {title} {\bibinfo {title} {Observation of two-dimensional plasmons and electron-ripplon scattering in a sheet of electrons on liquid helium},\ }\href {https://doi.org/10.1103/PhysRevLett.36.145} {\bibfield  {journal} {\bibinfo  {journal} {Phys. Rev. Lett}\ }\textbf {\bibinfo {volume} {36}},\ \bibinfo {pages} {145} (\bibinfo {year} {1976})}\BibitemShut {NoStop}%
\bibitem [{\citenamefont {Grimes}\ and\ \citenamefont {Adams}(1979)}]{grimes1979evidence}%
  \BibitemOpen
  \bibfield  {author} {\bibinfo {author} {\bibfnamefont {C.}~\bibnamefont {Grimes}}\ and\ \bibinfo {author} {\bibfnamefont {G.}~\bibnamefont {Adams}},\ }\bibfield  {title} {\bibinfo {title} {Evidence for a liquid-to-crystal phase transition in a classical, two-dimensional sheet of electrons},\ }\href {https://doi.org/10.1103/PhysRevLett.42.795} {\bibfield  {journal} {\bibinfo  {journal} {Phys. Rev. Lett}\ }\textbf {\bibinfo {volume} {42}},\ \bibinfo {pages} {795} (\bibinfo {year} {1979})}\BibitemShut {NoStop}%
\bibitem [{\citenamefont {Mast}\ \emph {et~al.}(1985)\citenamefont {Mast}, \citenamefont {Dahm},\ and\ \citenamefont {Fetter}}]{mast1985observation}%
  \BibitemOpen
  \bibfield  {author} {\bibinfo {author} {\bibfnamefont {D.}~\bibnamefont {Mast}}, \bibinfo {author} {\bibfnamefont {A.}~\bibnamefont {Dahm}},\ and\ \bibinfo {author} {\bibfnamefont {A.}~\bibnamefont {Fetter}},\ }\bibfield  {title} {\bibinfo {title} {Observation of bulk and edge magnetoplasmons in a two-dimensional electron fluid},\ }\href {https://doi.org/10.1103/PhysRevLett.54.1706} {\bibfield  {journal} {\bibinfo  {journal} {Phys. Rev. Lett}\ }\textbf {\bibinfo {volume} {54}},\ \bibinfo {pages} {1706} (\bibinfo {year} {1985})}\BibitemShut {NoStop}%
\bibitem [{\citenamefont {Glattli}\ \emph {et~al.}(1985)\citenamefont {Glattli}, \citenamefont {Andrei}, \citenamefont {Deville}, \citenamefont {Poitrenaud},\ and\ \citenamefont {Williams}}]{Glattli1985}%
  \BibitemOpen
  \bibfield  {author} {\bibinfo {author} {\bibfnamefont {D.}~\bibnamefont {Glattli}}, \bibinfo {author} {\bibfnamefont {E.}~\bibnamefont {Andrei}}, \bibinfo {author} {\bibfnamefont {G.}~\bibnamefont {Deville}}, \bibinfo {author} {\bibfnamefont {J.}~\bibnamefont {Poitrenaud}},\ and\ \bibinfo {author} {\bibfnamefont {F.}~\bibnamefont {Williams}},\ }\bibfield  {title} {\bibinfo {title} {Dynamical {Hall} effect in a two-dimensional classical plasma},\ }\href {https://doi.org/10.1103/PhysRevLett.54.1710} {\bibfield  {journal} {\bibinfo  {journal} {Phys. Rev. Lett.}\ }\textbf {\bibinfo {volume} {54}},\ \bibinfo {pages} {1710} (\bibinfo {year} {1985})}\BibitemShut {NoStop}%
\bibitem [{\citenamefont {Lea}\ \emph {et~al.}(1994)\citenamefont {Lea}, \citenamefont {Fozooni}, \citenamefont {Richardson},\ and\ \citenamefont {Blackburn}}]{Lea1994}%
  \BibitemOpen
  \bibfield  {author} {\bibinfo {author} {\bibfnamefont {M.}~\bibnamefont {Lea}}, \bibinfo {author} {\bibfnamefont {P.}~\bibnamefont {Fozooni}}, \bibinfo {author} {\bibfnamefont {P.}~\bibnamefont {Richardson}},\ and\ \bibinfo {author} {\bibfnamefont {A.}~\bibnamefont {Blackburn}},\ }\bibfield  {title} {\bibinfo {title} {Direct observation of many-electron magnetoconductivity in a nondegenerate 2d electron liquid},\ }\href {https://doi.org/10.1103/PhysRevLett.73.1142} {\bibfield  {journal} {\bibinfo  {journal} {Phys. Rev. Lett}\ }\textbf {\bibinfo {volume} {73}},\ \bibinfo {pages} {1142} (\bibinfo {year} {1994})}\BibitemShut {NoStop}%
\bibitem [{\citenamefont {Chepelianskii}\ \emph {et~al.}(2021)\citenamefont {Chepelianskii}, \citenamefont {Papoular}, \citenamefont {Konstantinov}, \citenamefont {Bouchiat},\ and\ \citenamefont {Kono}}]{Chepelianskii2021}%
  \BibitemOpen
  \bibfield  {author} {\bibinfo {author} {\bibfnamefont {A.}~\bibnamefont {Chepelianskii}}, \bibinfo {author} {\bibfnamefont {D.}~\bibnamefont {Papoular}}, \bibinfo {author} {\bibfnamefont {D.}~\bibnamefont {Konstantinov}}, \bibinfo {author} {\bibfnamefont {H.}~\bibnamefont {Bouchiat}},\ and\ \bibinfo {author} {\bibfnamefont {K.}~\bibnamefont {Kono}},\ }\bibfield  {title} {\bibinfo {title} {Coupled pair of one- and two-dimensional magnetoplasmons on electrons on helium},\ }\href {https://doi.org/10.1103/PhysRevB.103.075420} {\bibfield  {journal} {\bibinfo  {journal} {Phys. Rev. B}\ }\textbf {\bibinfo {volume} {103}},\ \bibinfo {pages} {075420} (\bibinfo {year} {2021})}\BibitemShut {NoStop}%
\bibitem [{\citenamefont {Kostylev}\ \emph {et~al.}(2024)\citenamefont {Kostylev}, \citenamefont {Hatifi}, \citenamefont {Konstantinov},\ and\ \citenamefont {Chepelianskii}}]{kostylev2024delocalized}%
  \BibitemOpen
  \bibfield  {author} {\bibinfo {author} {\bibfnamefont {I.}~\bibnamefont {Kostylev}}, \bibinfo {author} {\bibfnamefont {M.}~\bibnamefont {Hatifi}}, \bibinfo {author} {\bibfnamefont {D.}~\bibnamefont {Konstantinov}},\ and\ \bibinfo {author} {\bibfnamefont {A.}~\bibnamefont {Chepelianskii}},\ }\bibfield  {title} {\bibinfo {title} {Delocalized low-frequency magnetoplasmon in a two-dimensional electron fluid with cylindrical symmetry},\ }\href@noop {} {\bibfield  {journal} {\bibinfo  {journal} {arXiv preprint arXiv:2404.07582}\ } (\bibinfo {year} {2024})}\BibitemShut {NoStop}%
\bibitem [{\citenamefont {Yunusova}\ \emph {et~al.}(2019)\citenamefont {Yunusova}, \citenamefont {Konstantinov}, \citenamefont {Bouchiat},\ and\ \citenamefont {Chepelianskii}}]{yunusova2019coupling}%
  \BibitemOpen
  \bibfield  {author} {\bibinfo {author} {\bibfnamefont {K.~M.}\ \bibnamefont {Yunusova}}, \bibinfo {author} {\bibfnamefont {D.}~\bibnamefont {Konstantinov}}, \bibinfo {author} {\bibfnamefont {H.}~\bibnamefont {Bouchiat}},\ and\ \bibinfo {author} {\bibfnamefont {A.}~\bibnamefont {Chepelianskii}},\ }\bibfield  {title} {\bibinfo {title} {Coupling between {Rydberg} states and {Landau} levels of electrons trapped on liquid helium},\ }\href {https://doi.org/10.1103/PhysRevLett.122.176802} {\bibfield  {journal} {\bibinfo  {journal} {Phys. Rev. Lett}\ }\textbf {\bibinfo {volume} {122}},\ \bibinfo {pages} {176802} (\bibinfo {year} {2019})}\BibitemShut {NoStop}%
\bibitem [{\citenamefont {Zadorozhko}\ \emph {et~al.}(2021)\citenamefont {Zadorozhko}, \citenamefont {Chen}, \citenamefont {Chepelianskii},\ and\ \citenamefont {Konstantinov}}]{zadorozhko2021motional}%
  \BibitemOpen
  \bibfield  {author} {\bibinfo {author} {\bibfnamefont {A.}~\bibnamefont {Zadorozhko}}, \bibinfo {author} {\bibfnamefont {J.}~\bibnamefont {Chen}}, \bibinfo {author} {\bibfnamefont {A.}~\bibnamefont {Chepelianskii}},\ and\ \bibinfo {author} {\bibfnamefont {D.}~\bibnamefont {Konstantinov}},\ }\bibfield  {title} {\bibinfo {title} {Motional quantum states of surface electrons on liquid helium in a tilted magnetic field},\ }\href {https://doi.org/10.1103/PhysRevB.103.054507} {\bibfield  {journal} {\bibinfo  {journal} {Phys. Rev. B}\ }\textbf {\bibinfo {volume} {103}},\ \bibinfo {pages} {054507} (\bibinfo {year} {2021})}\BibitemShut {NoStop}%
\bibitem [{\citenamefont {Byeon}\ \emph {et~al.}(2021)\citenamefont {Byeon}, \citenamefont {Nasyedkin}, \citenamefont {Lane}, \citenamefont {Beysengulov}, \citenamefont {Zhang}, \citenamefont {Loloee},\ and\ \citenamefont {Pollanen}}]{byeon2021piezoacoustics}%
  \BibitemOpen
  \bibfield  {author} {\bibinfo {author} {\bibfnamefont {H.}~\bibnamefont {Byeon}}, \bibinfo {author} {\bibfnamefont {K.}~\bibnamefont {Nasyedkin}}, \bibinfo {author} {\bibfnamefont {J.}~\bibnamefont {Lane}}, \bibinfo {author} {\bibfnamefont {N.}~\bibnamefont {Beysengulov}}, \bibinfo {author} {\bibfnamefont {L.}~\bibnamefont {Zhang}}, \bibinfo {author} {\bibfnamefont {R.}~\bibnamefont {Loloee}},\ and\ \bibinfo {author} {\bibfnamefont {J.}~\bibnamefont {Pollanen}},\ }\bibfield  {title} {\bibinfo {title} {Piezoacoustics for precision control of electrons floating on helium},\ }\href {https://doi.org/10.1038/s41467-021-24452-7} {\bibfield  {journal} {\bibinfo  {journal} {Nat. Commun}\ }\textbf {\bibinfo {volume} {12}},\ \bibinfo {pages} {4150} (\bibinfo {year} {2021})}\BibitemShut {NoStop}%
\bibitem [{\citenamefont {Abdurakhimov}\ \emph {et~al.}(2016)\citenamefont {Abdurakhimov}, \citenamefont {Yamashiro}, \citenamefont {Badrutdinov},\ and\ \citenamefont {Konstantinov}}]{abdurakhimov2016strong}%
  \BibitemOpen
  \bibfield  {author} {\bibinfo {author} {\bibfnamefont {L.}~\bibnamefont {Abdurakhimov}}, \bibinfo {author} {\bibfnamefont {R.}~\bibnamefont {Yamashiro}}, \bibinfo {author} {\bibfnamefont {A.}~\bibnamefont {Badrutdinov}},\ and\ \bibinfo {author} {\bibfnamefont {D.}~\bibnamefont {Konstantinov}},\ }\bibfield  {title} {\bibinfo {title} {Strong coupling of the cyclotron motion of surface electrons on liquid helium to a microwave cavity},\ }\href {https://doi.org/10.1103/PhysRevLett.117.056803} {\bibfield  {journal} {\bibinfo  {journal} {Phys. Rev. Lett}\ }\textbf {\bibinfo {volume} {117}},\ \bibinfo {pages} {056803} (\bibinfo {year} {2016})}\BibitemShut {NoStop}%
\bibitem [{\citenamefont {Chen}\ \emph {et~al.}(2018)\citenamefont {Chen}, \citenamefont {Zadorozhko},\ and\ \citenamefont {Konstantinov}}]{chen2018strong}%
  \BibitemOpen
  \bibfield  {author} {\bibinfo {author} {\bibfnamefont {J.}~\bibnamefont {Chen}}, \bibinfo {author} {\bibfnamefont {A.}~\bibnamefont {Zadorozhko}},\ and\ \bibinfo {author} {\bibfnamefont {D.}~\bibnamefont {Konstantinov}},\ }\bibfield  {title} {\bibinfo {title} {Strong coupling of a two-dimensional electron ensemble to a single-mode cavity resonator},\ }\href {https://doi.org/10.1103/PhysRevB.98.235418} {\bibfield  {journal} {\bibinfo  {journal} {Phys. Rev. B}\ }\textbf {\bibinfo {volume} {98}},\ \bibinfo {pages} {235418} (\bibinfo {year} {2018})}\BibitemShut {NoStop}%
\bibitem [{\citenamefont {Yang}\ \emph {et~al.}(2016)\citenamefont {Yang}, \citenamefont {Fragner}, \citenamefont {Koolstra}, \citenamefont {Ocola}, \citenamefont {Czaplewski}, \citenamefont {Schoelkopf},\ and\ \citenamefont {Schuster}}]{yang2016coupling}%
  \BibitemOpen
  \bibfield  {author} {\bibinfo {author} {\bibfnamefont {G.}~\bibnamefont {Yang}}, \bibinfo {author} {\bibfnamefont {A.}~\bibnamefont {Fragner}}, \bibinfo {author} {\bibfnamefont {G.}~\bibnamefont {Koolstra}}, \bibinfo {author} {\bibfnamefont {L.}~\bibnamefont {Ocola}}, \bibinfo {author} {\bibfnamefont {D.}~\bibnamefont {Czaplewski}}, \bibinfo {author} {\bibfnamefont {R.}~\bibnamefont {Schoelkopf}},\ and\ \bibinfo {author} {\bibfnamefont {D.}~\bibnamefont {Schuster}},\ }\bibfield  {title} {\bibinfo {title} {Coupling an ensemble of electrons on superfluid helium to a superconducting circuit},\ }\href {https://doi.org/10.1103/PhysRevX.6.011031} {\bibfield  {journal} {\bibinfo  {journal} {Phys. Rev. X}\ }\textbf {\bibinfo {volume} {6}},\ \bibinfo {pages} {011031} (\bibinfo {year} {2016})}\BibitemShut {NoStop}%
\bibitem [{\citenamefont {Marty}(1986)}]{marty1986stability}%
  \BibitemOpen
  \bibfield  {author} {\bibinfo {author} {\bibfnamefont {D.}~\bibnamefont {Marty}},\ }\bibfield  {title} {\bibinfo {title} {Stability of two-dimensional electrons on a fractionated helium surface},\ }\href {https://doi.org/10.1088/0022-3719/19/30/019} {\bibfield  {journal} {\bibinfo  {journal} {J. Phys. C: Solid State Phys.}\ }\textbf {\bibinfo {volume} {19}},\ \bibinfo {pages} {6097} (\bibinfo {year} {1986})}\BibitemShut {NoStop}%
\bibitem [{\citenamefont {Rees}\ \emph {et~al.}(2011)\citenamefont {Rees}, \citenamefont {Kuroda}, \citenamefont {Marrache-Kikuchi}, \citenamefont {H{\"o}fer}, \citenamefont {Leiderer},\ and\ \citenamefont {Kono}}]{rees2011point}%
  \BibitemOpen
  \bibfield  {author} {\bibinfo {author} {\bibfnamefont {D.~G.}\ \bibnamefont {Rees}}, \bibinfo {author} {\bibfnamefont {I.}~\bibnamefont {Kuroda}}, \bibinfo {author} {\bibfnamefont {C.~A.}\ \bibnamefont {Marrache-Kikuchi}}, \bibinfo {author} {\bibfnamefont {M.}~\bibnamefont {H{\"o}fer}}, \bibinfo {author} {\bibfnamefont {P.}~\bibnamefont {Leiderer}},\ and\ \bibinfo {author} {\bibfnamefont {K.}~\bibnamefont {Kono}},\ }\bibfield  {title} {\bibinfo {title} {Point-contact transport properties of strongly correlated electrons on liquid helium},\ }\href {https://doi.org/10.1103/PhysRevLett.106.026803} {\bibfield  {journal} {\bibinfo  {journal} {Phys. Rev. Lett}\ }\textbf {\bibinfo {volume} {106}},\ \bibinfo {pages} {026803} (\bibinfo {year} {2011})}\BibitemShut {NoStop}%
\bibitem [{\citenamefont {Rees}\ \emph {et~al.}(2016{\natexlab{a}})\citenamefont {Rees}, \citenamefont {Beysengulov}, \citenamefont {Lin},\ and\ \citenamefont {Kono}}]{rees2016stick}%
  \BibitemOpen
  \bibfield  {author} {\bibinfo {author} {\bibfnamefont {D.~G.}\ \bibnamefont {Rees}}, \bibinfo {author} {\bibfnamefont {N.~R.}\ \bibnamefont {Beysengulov}}, \bibinfo {author} {\bibfnamefont {J.-J.}\ \bibnamefont {Lin}},\ and\ \bibinfo {author} {\bibfnamefont {K.}~\bibnamefont {Kono}},\ }\bibfield  {title} {\bibinfo {title} {Stick-slip motion of the {Wigner} solid on liquid helium},\ }\href {https://doi.org/10.1103/PhysRevLett.116.206801} {\bibfield  {journal} {\bibinfo  {journal} {Phys. Rev. Lett}\ }\textbf {\bibinfo {volume} {116}},\ \bibinfo {pages} {206801} (\bibinfo {year} {2016}{\natexlab{a}})}\BibitemShut {NoStop}%
\bibitem [{\citenamefont {Zou}\ \emph {et~al.}(2021)\citenamefont {Zou}, \citenamefont {Konstantinov},\ and\ \citenamefont {Rees}}]{ZouPRB2021}%
  \BibitemOpen
  \bibfield  {author} {\bibinfo {author} {\bibfnamefont {S.}~\bibnamefont {Zou}}, \bibinfo {author} {\bibfnamefont {D.}~\bibnamefont {Konstantinov}},\ and\ \bibinfo {author} {\bibfnamefont {D.~G.}\ \bibnamefont {Rees}},\ }\bibfield  {title} {\bibinfo {title} {Dynamical ordering in a two-dimensional electron crystal confined in a narrow channel geometry},\ }\href {https://doi.org/10.1103/PhysRevB.104.045427} {\bibfield  {journal} {\bibinfo  {journal} {Phys. Rev. B}\ }\textbf {\bibinfo {volume} {104}},\ \bibinfo {pages} {045427} (\bibinfo {year} {2021})}\BibitemShut {NoStop}%
\bibitem [{\citenamefont {Glasson}\ \emph {et~al.}(2001)\citenamefont {Glasson}, \citenamefont {Dotsenko}, \citenamefont {Fozooni}, \citenamefont {Lea}, \citenamefont {Bailey}, \citenamefont {Papageorgiou}, \citenamefont {Andresen},\ and\ \citenamefont {Kristensen}}]{glasson2001observation}%
  \BibitemOpen
  \bibfield  {author} {\bibinfo {author} {\bibfnamefont {P.}~\bibnamefont {Glasson}}, \bibinfo {author} {\bibfnamefont {V.}~\bibnamefont {Dotsenko}}, \bibinfo {author} {\bibfnamefont {P.}~\bibnamefont {Fozooni}}, \bibinfo {author} {\bibfnamefont {M.}~\bibnamefont {Lea}}, \bibinfo {author} {\bibfnamefont {W.}~\bibnamefont {Bailey}}, \bibinfo {author} {\bibfnamefont {G.}~\bibnamefont {Papageorgiou}}, \bibinfo {author} {\bibfnamefont {S.}~\bibnamefont {Andresen}},\ and\ \bibinfo {author} {\bibfnamefont {A.}~\bibnamefont {Kristensen}},\ }\bibfield  {title} {\bibinfo {title} {Observation of dynamical ordering in a confined {Wigner} crystal},\ }\href {https://doi.org/10.1103/PhysRevLett.87.176802} {\bibfield  {journal} {\bibinfo  {journal} {Phys. Rev. Lett}\ }\textbf {\bibinfo {volume} {87}},\ \bibinfo {pages} {176802} (\bibinfo {year} {2001})}\BibitemShut {NoStop}%
\bibitem [{\citenamefont {Rees}\ \emph {et~al.}(2016{\natexlab{b}})\citenamefont {Rees}, \citenamefont {Beysengulov}, \citenamefont {Teranishi}, \citenamefont {Tsao}, \citenamefont {Yeh}, \citenamefont {Chiu}, \citenamefont {Lin}, \citenamefont {Tayurskii}, \citenamefont {Lin},\ and\ \citenamefont {Kono}}]{rees2016structural}%
  \BibitemOpen
  \bibfield  {author} {\bibinfo {author} {\bibfnamefont {D.~G.}\ \bibnamefont {Rees}}, \bibinfo {author} {\bibfnamefont {N.~R.}\ \bibnamefont {Beysengulov}}, \bibinfo {author} {\bibfnamefont {Y.}~\bibnamefont {Teranishi}}, \bibinfo {author} {\bibfnamefont {C.-S.}\ \bibnamefont {Tsao}}, \bibinfo {author} {\bibfnamefont {S.-S.}\ \bibnamefont {Yeh}}, \bibinfo {author} {\bibfnamefont {S.-P.}\ \bibnamefont {Chiu}}, \bibinfo {author} {\bibfnamefont {Y.-H.}\ \bibnamefont {Lin}}, \bibinfo {author} {\bibfnamefont {D.~A.}\ \bibnamefont {Tayurskii}}, \bibinfo {author} {\bibfnamefont {J.-J.}\ \bibnamefont {Lin}},\ and\ \bibinfo {author} {\bibfnamefont {K.}~\bibnamefont {Kono}},\ }\bibfield  {title} {\bibinfo {title} {Structural order and melting of a quasi-one-dimensional electron system},\ }\href {https://doi.org/10.1103/PhysRevB.94.045139} {\bibfield  {journal} {\bibinfo  {journal} {Phys. Rev. B}\ }\textbf {\bibinfo {volume} {94}},\ \bibinfo {pages} {045139} (\bibinfo {year} {2016}{\natexlab{b}})}\BibitemShut {NoStop}%
\bibitem [{\citenamefont {Sabouret}\ \emph {et~al.}(2008)\citenamefont {Sabouret}, \citenamefont {Bradbury}, \citenamefont {Shankar}, \citenamefont {Bert},\ and\ \citenamefont {Lyon}}]{sabouret2008signal}%
  \BibitemOpen
  \bibfield  {author} {\bibinfo {author} {\bibfnamefont {G.}~\bibnamefont {Sabouret}}, \bibinfo {author} {\bibfnamefont {F.}~\bibnamefont {Bradbury}}, \bibinfo {author} {\bibfnamefont {S.}~\bibnamefont {Shankar}}, \bibinfo {author} {\bibfnamefont {J.}~\bibnamefont {Bert}},\ and\ \bibinfo {author} {\bibfnamefont {S.}~\bibnamefont {Lyon}},\ }\bibfield  {title} {\bibinfo {title} {Signal and charge transfer efficiency of few electrons clocked on microscopic superfluid helium channels},\ }\href {https://doi.org/10.1063/1.2884693} {\bibfield  {journal} {\bibinfo  {journal} {Appl. Phys. Lett}\ }\textbf {\bibinfo {volume} {92}},\ \bibinfo {pages} {082104} (\bibinfo {year} {2008})}\BibitemShut {NoStop}%
\bibitem [{\citenamefont {Bradbury}\ \emph {et~al.}(2011)\citenamefont {Bradbury}, \citenamefont {Takita}, \citenamefont {Gurrieri}, \citenamefont {Wilkel}, \citenamefont {Eng}, \citenamefont {Carroll},\ and\ \citenamefont {Lyon}}]{bradbury2011efficient}%
  \BibitemOpen
  \bibfield  {author} {\bibinfo {author} {\bibfnamefont {F.}~\bibnamefont {Bradbury}}, \bibinfo {author} {\bibfnamefont {M.}~\bibnamefont {Takita}}, \bibinfo {author} {\bibfnamefont {T.}~\bibnamefont {Gurrieri}}, \bibinfo {author} {\bibfnamefont {K.}~\bibnamefont {Wilkel}}, \bibinfo {author} {\bibfnamefont {K.}~\bibnamefont {Eng}}, \bibinfo {author} {\bibfnamefont {M.}~\bibnamefont {Carroll}},\ and\ \bibinfo {author} {\bibfnamefont {S.~A.}\ \bibnamefont {Lyon}},\ }\bibfield  {title} {\bibinfo {title} {Efficient clocked electron transfer on superfluid helium},\ }\href {https://doi.org/10.1103/PhysRevLett.107.266803} {\bibfield  {journal} {\bibinfo  {journal} {Phys. Rev. Lett}\ }\textbf {\bibinfo {volume} {107}},\ \bibinfo {pages} {266803} (\bibinfo {year} {2011})}\BibitemShut {NoStop}%
\bibitem [{\citenamefont {Beysengulov}\ \emph {et~al.}(2016)\citenamefont {Beysengulov}, \citenamefont {Rees}, \citenamefont {Lysogorskiy}, \citenamefont {Galiullin}, \citenamefont {Vazjukov}, \citenamefont {Tayurskii},\ and\ \citenamefont {Kono}}]{beysengulov2016structural}%
  \BibitemOpen
  \bibfield  {author} {\bibinfo {author} {\bibfnamefont {N.}~\bibnamefont {Beysengulov}}, \bibinfo {author} {\bibfnamefont {D.}~\bibnamefont {Rees}}, \bibinfo {author} {\bibfnamefont {Y.}~\bibnamefont {Lysogorskiy}}, \bibinfo {author} {\bibfnamefont {N.}~\bibnamefont {Galiullin}}, \bibinfo {author} {\bibfnamefont {A.}~\bibnamefont {Vazjukov}}, \bibinfo {author} {\bibfnamefont {D.}~\bibnamefont {Tayurskii}},\ and\ \bibinfo {author} {\bibfnamefont {K.}~\bibnamefont {Kono}},\ }\bibfield  {title} {\bibinfo {title} {Structural transitions in a quasi-1d {Wigner} solid on liquid helium},\ }\href {https://doi.org/10.1007/s10909-015-1344-4} {\bibfield  {journal} {\bibinfo  {journal} {J. Low Temp. Phys}\ }\textbf {\bibinfo {volume} {182}},\ \bibinfo {pages} {28} (\bibinfo {year} {2016})}\BibitemShut {NoStop}%
\bibitem [{\citenamefont {Monarkha}\ and\ \citenamefont {Kono}(2004)}]{monarkha2004two}%
  \BibitemOpen
  \bibfield  {author} {\bibinfo {author} {\bibfnamefont {Y.}~\bibnamefont {Monarkha}}\ and\ \bibinfo {author} {\bibfnamefont {K.}~\bibnamefont {Kono}},\ }\href {https://doi.org/10.1007/978-3-662-10639-6} {\emph {\bibinfo {title} {Two-Dimensional Coulomb Liquids and Solids}}}\ (\bibinfo  {publisher} {Springer Berlin, Heidelberg},\ \bibinfo {year} {2004})\BibitemShut {NoStop}%
\bibitem [{\citenamefont {Andrei}(1997)}]{andrei1997two}%
  \BibitemOpen
  \bibinfo {editor} {\bibfnamefont {E.~Y.}\ \bibnamefont {Andrei}},\ ed.,\ \href {https://doi.org/10.1007/978-94-015-1286-2} {\emph {\bibinfo {title} {Two-Dimensional Electron Systems: on Helium and other Cryogenic Substrates}}}\ (\bibinfo  {publisher} {Springer Dordrecht},\ \bibinfo {year} {1997})\BibitemShut {NoStop}%
\bibitem [{\citenamefont {Dykman}\ and\ \citenamefont {Rubo}(1997)}]{dykman1997bragg}%
  \BibitemOpen
  \bibfield  {author} {\bibinfo {author} {\bibfnamefont {M.}~\bibnamefont {Dykman}}\ and\ \bibinfo {author} {\bibfnamefont {Y.~G.}\ \bibnamefont {Rubo}},\ }\bibfield  {title} {\bibinfo {title} {Bragg-cherenkov scattering and nonlinear conductivity of a two-dimensional {Wigner} crystal},\ }\href {https://doi.org/10.1103/PhysRevLett.78.4813} {\bibfield  {journal} {\bibinfo  {journal} {Phys. Rev. Lett}\ }\textbf {\bibinfo {volume} {78}},\ \bibinfo {pages} {4813} (\bibinfo {year} {1997})}\BibitemShut {NoStop}%
\bibitem [{\citenamefont {Rees}\ \emph {et~al.}(2012)\citenamefont {Rees}, \citenamefont {Totsuji},\ and\ \citenamefont {Kono}}]{rees2012}%
  \BibitemOpen
  \bibfield  {author} {\bibinfo {author} {\bibfnamefont {D.~G.}\ \bibnamefont {Rees}}, \bibinfo {author} {\bibfnamefont {H.}~\bibnamefont {Totsuji}},\ and\ \bibinfo {author} {\bibfnamefont {K.}~\bibnamefont {Kono}},\ }\bibfield  {title} {\bibinfo {title} {Commensurability-dependent transport of a {Wigner} crystal in a nanoconstriction},\ }\href {https://doi.org/10.1103/PhysRevLett.108.176801} {\bibfield  {journal} {\bibinfo  {journal} {Phys. Rev. Lett}\ }\textbf {\bibinfo {volume} {108}},\ \bibinfo {pages} {176801} (\bibinfo {year} {2012})}\BibitemShut {NoStop}%
\bibitem [{\citenamefont {Rees}\ \emph {et~al.}(2020)\citenamefont {Rees}, \citenamefont {Yeh}, \citenamefont {Lee}, \citenamefont {Schnyder}, \citenamefont {Williams}, \citenamefont {Lin},\ and\ \citenamefont {Kono}}]{rees2020dynamical}%
  \BibitemOpen
  \bibfield  {author} {\bibinfo {author} {\bibfnamefont {D.~G.}\ \bibnamefont {Rees}}, \bibinfo {author} {\bibfnamefont {S.-S.}\ \bibnamefont {Yeh}}, \bibinfo {author} {\bibfnamefont {B.-C.}\ \bibnamefont {Lee}}, \bibinfo {author} {\bibfnamefont {S.~K.}\ \bibnamefont {Schnyder}}, \bibinfo {author} {\bibfnamefont {F.~I.}\ \bibnamefont {Williams}}, \bibinfo {author} {\bibfnamefont {J.-J.}\ \bibnamefont {Lin}},\ and\ \bibinfo {author} {\bibfnamefont {K.}~\bibnamefont {Kono}},\ }\bibfield  {title} {\bibinfo {title} {Dynamical decoupling and recoupling of the {Wigner} solid to a liquid helium substrate},\ }\href {https://doi.org/10.1103/PhysRevB.102.075439} {\bibfield  {journal} {\bibinfo  {journal} {Phys. Rev. B}\ }\textbf {\bibinfo {volume} {102}},\ \bibinfo {pages} {075439} (\bibinfo {year} {2020})}\BibitemShut {NoStop}%
\bibitem [{\citenamefont {Konstantinov}\ \emph {et~al.}(2008)\citenamefont {Konstantinov}, \citenamefont {Isshiki}, \citenamefont {Akimoto}, \citenamefont {Shirahama}, \citenamefont {Monarkha},\ and\ \citenamefont {Kono}}]{konstantinov2008microwave}%
  \BibitemOpen
  \bibfield  {author} {\bibinfo {author} {\bibfnamefont {D.}~\bibnamefont {Konstantinov}}, \bibinfo {author} {\bibfnamefont {H.}~\bibnamefont {Isshiki}}, \bibinfo {author} {\bibfnamefont {H.}~\bibnamefont {Akimoto}}, \bibinfo {author} {\bibfnamefont {K.}~\bibnamefont {Shirahama}}, \bibinfo {author} {\bibfnamefont {Y.}~\bibnamefont {Monarkha}},\ and\ \bibinfo {author} {\bibfnamefont {K.}~\bibnamefont {Kono}},\ }\bibfield  {title} {\bibinfo {title} {Microwave-absorption-induced heating of surface state electrons on liquid 3{He}},\ }\href {https://doi.org/10.1143/JPSJ.77.034705} {\bibfield  {journal} {\bibinfo  {journal} {J. Phys. Soc. Jpn}\ }\textbf {\bibinfo {volume} {77}},\ \bibinfo {pages} {034705} (\bibinfo {year} {2008})}\BibitemShut {NoStop}%
\bibitem [{\citenamefont {Nasyedkin}\ \emph {et~al.}(2011)\citenamefont {Nasyedkin}, \citenamefont {Syvokon},\ and\ \citenamefont {Monarkha}}]{nasyedkin2011transport}%
  \BibitemOpen
  \bibfield  {author} {\bibinfo {author} {\bibfnamefont {K.}~\bibnamefont {Nasyedkin}}, \bibinfo {author} {\bibfnamefont {V.}~\bibnamefont {Syvokon}},\ and\ \bibinfo {author} {\bibfnamefont {Y.}~\bibnamefont {Monarkha}},\ }\bibfield  {title} {\bibinfo {title} {Transport properties of the two-dimensional {Wigner} solid on liquid helium in the presence of a high-frequency damaging electric field},\ }\href {https://doi.org/10.1007/s10909-010-0338-5} {\bibfield  {journal} {\bibinfo  {journal} {J. Low Temp. Phys}\ }\textbf {\bibinfo {volume} {163}},\ \bibinfo {pages} {148} (\bibinfo {year} {2011})}\BibitemShut {NoStop}%
\bibitem [{\citenamefont {Zou}\ \emph {et~al.}(2022)\citenamefont {Zou}, \citenamefont {Grossenbach},\ and\ \citenamefont {Konstantinov}}]{zou2022observation}%
  \BibitemOpen
  \bibfield  {author} {\bibinfo {author} {\bibfnamefont {S.}~\bibnamefont {Zou}}, \bibinfo {author} {\bibfnamefont {S.}~\bibnamefont {Grossenbach}},\ and\ \bibinfo {author} {\bibfnamefont {D.}~\bibnamefont {Konstantinov}},\ }\bibfield  {title} {\bibinfo {title} {Observation of the {Rydberg} resonance in surface electrons on superfluid helium confined in a 4-$\mu$ m deep channel},\ }\href {https://doi.org/10.1007/s10909-022-02749-1} {\bibfield  {journal} {\bibinfo  {journal} {J. Low Temp. Phys}\ }\textbf {\bibinfo {volume} {208}},\ \bibinfo {pages} {211} (\bibinfo {year} {2022})}\BibitemShut {NoStop}%
\bibitem [{\citenamefont {Saitoh}(1977)}]{saitoh1977warm}%
  \BibitemOpen
  \bibfield  {author} {\bibinfo {author} {\bibfnamefont {M.}~\bibnamefont {Saitoh}},\ }\bibfield  {title} {\bibinfo {title} {Warm electrons on the liquid 4{He} surface},\ }\href {https://doi.org/10.1143/JPSJ.42.201} {\bibfield  {journal} {\bibinfo  {journal} {J. Phys. Soc. Jpn}\ }\textbf {\bibinfo {volume} {42}},\ \bibinfo {pages} {201} (\bibinfo {year} {1977})}\BibitemShut {NoStop}%
\bibitem [{\citenamefont {Giannetta}\ and\ \citenamefont {Wilen}(1991)}]{giannetta1991}%
  \BibitemOpen
  \bibfield  {author} {\bibinfo {author} {\bibfnamefont {R.}~\bibnamefont {Giannetta}}\ and\ \bibinfo {author} {\bibfnamefont {L.}~\bibnamefont {Wilen}},\ }\bibfield  {title} {\bibinfo {title} {Nonequilibrium melting of the two dimensional electron crystal},\ }\href {https://doi.org/https://doi.org/10.1016/0038-1098(91)90283-2} {\bibfield  {journal} {\bibinfo  {journal} {Solid State Commun}\ }\textbf {\bibinfo {volume} {78}},\ \bibinfo {pages} {199} (\bibinfo {year} {1991})}\BibitemShut {NoStop}%
\bibitem [{\citenamefont {Shirahama}\ and\ \citenamefont {Kono}(1995)}]{shirahama1995dynamical}%
  \BibitemOpen
  \bibfield  {author} {\bibinfo {author} {\bibfnamefont {K.}~\bibnamefont {Shirahama}}\ and\ \bibinfo {author} {\bibfnamefont {K.}~\bibnamefont {Kono}},\ }\bibfield  {title} {\bibinfo {title} {Dynamical transition in the {Wigner} solid on a liquid helium surface},\ }\href {https://doi.org/10.1103/PhysRevLett.74.781} {\bibfield  {journal} {\bibinfo  {journal} {Phys. Rev. Lett}\ }\textbf {\bibinfo {volume} {74}},\ \bibinfo {pages} {781} (\bibinfo {year} {1995})}\BibitemShut {NoStop}%
\bibitem [{\citenamefont {Ikegami}\ \emph {et~al.}(2009)\citenamefont {Ikegami}, \citenamefont {Akimoto},\ and\ \citenamefont {Kono}}]{ikegami2009nonlinear}%
  \BibitemOpen
  \bibfield  {author} {\bibinfo {author} {\bibfnamefont {H.}~\bibnamefont {Ikegami}}, \bibinfo {author} {\bibfnamefont {H.}~\bibnamefont {Akimoto}},\ and\ \bibinfo {author} {\bibfnamefont {K.}~\bibnamefont {Kono}},\ }\bibfield  {title} {\bibinfo {title} {Nonlinear transport of the {Wigner} solid on superfluid {He}4 in a channel geometry},\ }\href {https://doi.org/10.1103/PhysRevLett.102.046807} {\bibfield  {journal} {\bibinfo  {journal} {Phys. Rev. Lett}\ }\textbf {\bibinfo {volume} {102}},\ \bibinfo {pages} {046807} (\bibinfo {year} {2009})}\BibitemShut {NoStop}%
\bibitem [{\citenamefont {Glattli}\ \emph {et~al.}(1988)\citenamefont {Glattli}, \citenamefont {Andrei},\ and\ \citenamefont {Williams}}]{glattli1988thermodynamic}%
  \BibitemOpen
  \bibfield  {author} {\bibinfo {author} {\bibfnamefont {D.}~\bibnamefont {Glattli}}, \bibinfo {author} {\bibfnamefont {E.}~\bibnamefont {Andrei}},\ and\ \bibinfo {author} {\bibfnamefont {F.}~\bibnamefont {Williams}},\ }\bibfield  {title} {\bibinfo {title} {Thermodynamic measurement on the melting of a two-dimensional electron solid},\ }\href {https://doi.org/10.1103/PhysRevLett.60.420} {\bibfield  {journal} {\bibinfo  {journal} {Phys. Rev. Lett}\ }\textbf {\bibinfo {volume} {60}},\ \bibinfo {pages} {420} (\bibinfo {year} {1988})}\BibitemShut {NoStop}%
\bibitem [{\citenamefont {Dykman}\ \emph {et~al.}(2003)\citenamefont {Dykman}, \citenamefont {Platzman},\ and\ \citenamefont {Seddighrad}}]{dykman2003}%
  \BibitemOpen
  \bibfield  {author} {\bibinfo {author} {\bibfnamefont {M.~I.}\ \bibnamefont {Dykman}}, \bibinfo {author} {\bibfnamefont {P.~M.}\ \bibnamefont {Platzman}},\ and\ \bibinfo {author} {\bibfnamefont {P.}~\bibnamefont {Seddighrad}},\ }\bibfield  {title} {\bibinfo {title} {Qubits with electrons on liquid helium},\ }\href {https://doi.org/10.1103/PhysRevB.67.155402} {\bibfield  {journal} {\bibinfo  {journal} {Phys. Rev. B}\ }\textbf {\bibinfo {volume} {67}},\ \bibinfo {pages} {155402} (\bibinfo {year} {2003})}\BibitemShut {NoStop}%
\bibitem [{\citenamefont {Fisher}\ \emph {et~al.}(1979)\citenamefont {Fisher}, \citenamefont {Halperin},\ and\ \citenamefont {Platzman}}]{fisher1979phonon}%
  \BibitemOpen
  \bibfield  {author} {\bibinfo {author} {\bibfnamefont {D.~S.}\ \bibnamefont {Fisher}}, \bibinfo {author} {\bibfnamefont {B.}~\bibnamefont {Halperin}},\ and\ \bibinfo {author} {\bibfnamefont {P.}~\bibnamefont {Platzman}},\ }\bibfield  {title} {\bibinfo {title} {Phonon-ripplon coupling and the two-dimensional electron solid on a liquid-helium surface},\ }\href {https://doi.org/10.1103/PhysRevLett.42.798} {\bibfield  {journal} {\bibinfo  {journal} {Phys. Rev. Lett}\ }\textbf {\bibinfo {volume} {42}},\ \bibinfo {pages} {798} (\bibinfo {year} {1979})}\BibitemShut {NoStop}%
\bibitem [{\citenamefont {Stan}\ and\ \citenamefont {Dahm}(1989)}]{StanDahm1989}%
  \BibitemOpen
  \bibfield  {author} {\bibinfo {author} {\bibfnamefont {M.~A.}\ \bibnamefont {Stan}}\ and\ \bibinfo {author} {\bibfnamefont {A.~J.}\ \bibnamefont {Dahm}},\ }\bibfield  {title} {\bibinfo {title} {Two-dimensional melting: Electrons on helium},\ }\href {https://doi.org/10.1103/PhysRevB.40.8995} {\bibfield  {journal} {\bibinfo  {journal} {Phys. Rev. B}\ }\textbf {\bibinfo {volume} {40}},\ \bibinfo {pages} {8995} (\bibinfo {year} {1989})}\BibitemShut {NoStop}%
\bibitem [{\citenamefont {Konstantinov}\ \emph {et~al.}(2012)\citenamefont {Konstantinov}, \citenamefont {Chepelianskii},\ and\ \citenamefont {Kono}}]{Konstantinov2012}%
  \BibitemOpen
  \bibfield  {author} {\bibinfo {author} {\bibfnamefont {D.}~\bibnamefont {Konstantinov}}, \bibinfo {author} {\bibfnamefont {A.}~\bibnamefont {Chepelianskii}},\ and\ \bibinfo {author} {\bibfnamefont {K.}~\bibnamefont {Kono}},\ }\bibfield  {title} {\bibinfo {title} {Resonant photovoltaic effect in surface state electrons on liquid helium},\ }\href {https://doi.org/10.1143/JPSJ.81.093601} {\bibfield  {journal} {\bibinfo  {journal} {J. Phys. Soc. Jpn}\ }\textbf {\bibinfo {volume} {81}},\ \bibinfo {pages} {093601} (\bibinfo {year} {2012})}\BibitemShut {NoStop}%
\bibitem [{\citenamefont {Kapralov}\ and\ \citenamefont {Svintsov}(2020)}]{Kapralov2020}%
  \BibitemOpen
  \bibfield  {author} {\bibinfo {author} {\bibfnamefont {K.}~\bibnamefont {Kapralov}}\ and\ \bibinfo {author} {\bibfnamefont {D.}~\bibnamefont {Svintsov}},\ }\bibfield  {title} {\bibinfo {title} {Plasmon damping in electronically open systems},\ }\href {https://doi.org/10.1103/PhysRevLett.125.236801} {\bibfield  {journal} {\bibinfo  {journal} {Phys. Rev. Lett.}\ }\textbf {\bibinfo {volume} {125}},\ \bibinfo {pages} {236801} (\bibinfo {year} {2020})}\BibitemShut {NoStop}%
\bibitem [{\citenamefont {Satou}\ \emph {et~al.}(2005)\citenamefont {Satou}, \citenamefont {Ryzhii},\ and\ \citenamefont {Chaplik}}]{Satou2005}%
  \BibitemOpen
  \bibfield  {author} {\bibinfo {author} {\bibfnamefont {A.}~\bibnamefont {Satou}}, \bibinfo {author} {\bibfnamefont {V.}~\bibnamefont {Ryzhii}},\ and\ \bibinfo {author} {\bibfnamefont {A.}~\bibnamefont {Chaplik}},\ }\bibfield  {title} {\bibinfo {title} {Plasma oscillations in two-dimensional electron channel with nonideally conducting side contacts},\ }\href {https://doi.org/10.1063/1.1993756} {\bibfield  {journal} {\bibinfo  {journal} {J. Appl. Phys}\ }\textbf {\bibinfo {volume} {98}},\ \bibinfo {pages} {034502} (\bibinfo {year} {2005})}\BibitemShut {NoStop}%
\bibitem [{\citenamefont {Torre}\ \emph {et~al.}(2019)\citenamefont {Torre}, \citenamefont {Vieira~de Castro}, \citenamefont {Van~Duppen}, \citenamefont {Barcons~Ruiz}, \citenamefont {Peeters}, \citenamefont {Koppens},\ and\ \citenamefont {Polini}}]{Torre2019}%
  \BibitemOpen
  \bibfield  {author} {\bibinfo {author} {\bibfnamefont {I.}~\bibnamefont {Torre}}, \bibinfo {author} {\bibfnamefont {L.}~\bibnamefont {Vieira~de Castro}}, \bibinfo {author} {\bibfnamefont {B.}~\bibnamefont {Van~Duppen}}, \bibinfo {author} {\bibfnamefont {D.}~\bibnamefont {Barcons~Ruiz}}, \bibinfo {author} {\bibfnamefont {F.}~\bibnamefont {Peeters}}, \bibinfo {author} {\bibfnamefont {F.~H.~L.}\ \bibnamefont {Koppens}},\ and\ \bibinfo {author} {\bibfnamefont {M.}~\bibnamefont {Polini}},\ }\bibfield  {title} {\bibinfo {title} {Acoustic plasmons at the crossover between the collisionless and hydrodynamic regimes in two-dimensional electron liquids},\ }\href {https://doi.org/10.1103/PhysRevB.99.144307} {\bibfield  {journal} {\bibinfo  {journal} {Phys. Rev. B}\ }\textbf {\bibinfo {volume} {99}},\ \bibinfo {pages} {144307} (\bibinfo {year} {2019})}\BibitemShut {NoStop}%
\bibitem [{\citenamefont {Serniak}\ \emph {et~al.}(2019)\citenamefont {Serniak}, \citenamefont {Diamond}, \citenamefont {Hays}, \citenamefont {Fatemi}, \citenamefont {Shankar}, \citenamefont {Frunzio}, \citenamefont {Schoelkopf},\ and\ \citenamefont {Devoret}}]{serniak2019direct}%
  \BibitemOpen
  \bibfield  {author} {\bibinfo {author} {\bibfnamefont {K.}~\bibnamefont {Serniak}}, \bibinfo {author} {\bibfnamefont {S.}~\bibnamefont {Diamond}}, \bibinfo {author} {\bibfnamefont {M.}~\bibnamefont {Hays}}, \bibinfo {author} {\bibfnamefont {V.}~\bibnamefont {Fatemi}}, \bibinfo {author} {\bibfnamefont {S.}~\bibnamefont {Shankar}}, \bibinfo {author} {\bibfnamefont {L.}~\bibnamefont {Frunzio}}, \bibinfo {author} {\bibfnamefont {R.}~\bibnamefont {Schoelkopf}},\ and\ \bibinfo {author} {\bibfnamefont {M.}~\bibnamefont {Devoret}},\ }\bibfield  {title} {\bibinfo {title} {Direct dispersive monitoring of charge parity in offset-charge-sensitive transmons},\ }\href {https://doi.org/10.1103/PhysRevApplied.12.014052} {\bibfield  {journal} {\bibinfo  {journal} {Phys. Rev. A}\ }\textbf {\bibinfo {volume} {12}},\ \bibinfo {pages} {014052} (\bibinfo {year} {2019})}\BibitemShut {NoStop}%
\bibitem [{\citenamefont {Dykman}(2016)}]{Dykman2016Phys}%
  \BibitemOpen
  \bibfield  {author} {\bibinfo {author} {\bibfnamefont {M.~I.}\ \bibnamefont {Dykman}},\ }\bibfield  {title} {\bibinfo {title} {Stick-slip motion in a quantum field},\ }\href {https://physics.aps.org/articles/v9/54} {\bibfield  {journal} {\bibinfo  {journal} {Physics}\ }\textbf {\bibinfo {volume} {9}},\ \bibinfo {pages} {54} (\bibinfo {year} {2016})}\BibitemShut {NoStop}%
\bibitem [{\citenamefont {Martinis}\ \emph {et~al.}(2005)\citenamefont {Martinis}, \citenamefont {Cooper}, \citenamefont {McDermott}, \citenamefont {Steffen}, \citenamefont {Ansmann}, \citenamefont {Osborn}, \citenamefont {Cicak}, \citenamefont {Oh}, \citenamefont {Pappas}, \citenamefont {Simmonds},\ and\ \citenamefont {Yu}}]{martinis2005decoherence}%
  \BibitemOpen
  \bibfield  {author} {\bibinfo {author} {\bibfnamefont {J.~M.}\ \bibnamefont {Martinis}}, \bibinfo {author} {\bibfnamefont {K.}~\bibnamefont {Cooper}}, \bibinfo {author} {\bibfnamefont {R.}~\bibnamefont {McDermott}}, \bibinfo {author} {\bibfnamefont {M.}~\bibnamefont {Steffen}}, \bibinfo {author} {\bibfnamefont {M.}~\bibnamefont {Ansmann}}, \bibinfo {author} {\bibfnamefont {K.}~\bibnamefont {Osborn}}, \bibinfo {author} {\bibfnamefont {K.}~\bibnamefont {Cicak}}, \bibinfo {author} {\bibfnamefont {S.}~\bibnamefont {Oh}}, \bibinfo {author} {\bibfnamefont {D.}~\bibnamefont {Pappas}}, \bibinfo {author} {\bibfnamefont {R.}~\bibnamefont {Simmonds}},\ and\ \bibinfo {author} {\bibfnamefont {C.~C.}\ \bibnamefont {Yu}},\ }\bibfield  {title} {\bibinfo {title} {Decoherence in {Josephson} qubits from dielectric loss},\ }\href {https://doi.org/10.1103/PhysRevLett.95.210503} {\bibfield  {journal} {\bibinfo  {journal} {Phys. Rev. Lett.}\ }\textbf {\bibinfo {volume} {95}},\ \bibinfo {pages} {210503} (\bibinfo {year}
  {2005})}\BibitemShut {NoStop}%
\bibitem [{\citenamefont {Klimov}\ \emph {et~al.}(2018)\citenamefont {Klimov}, \citenamefont {Kelly}, \citenamefont {Chen}, \citenamefont {Neeley}, \citenamefont {Megrant}, \citenamefont {Burkett}, \citenamefont {Barends}, \citenamefont {Arya}, \citenamefont {Chiaro}, \citenamefont {Chen}, \citenamefont {Dunsworth}, \citenamefont {Fowler}, \citenamefont {Foxen}, \citenamefont {Gidney}, \citenamefont {Giustina}, \citenamefont {Graff}, \citenamefont {Huang}, \citenamefont {Jeffrey}, \citenamefont {Lucero}, \citenamefont {Mutus}, \citenamefont {Naaman}, \citenamefont {Neill}, \citenamefont {Quintana}, \citenamefont {Roushan}, \citenamefont {Sank}, \citenamefont {Vainsencher}, \citenamefont {Wenner}, \citenamefont {White}, \citenamefont {Boixo}, \citenamefont {Babbush}, \citenamefont {Smelyanskiy}, \citenamefont {Neven},\ and\ \citenamefont {Martinis}}]{klimov2018fluctuations}%
  \BibitemOpen
  \bibfield  {author} {\bibinfo {author} {\bibfnamefont {P.}~\bibnamefont {Klimov}}, \bibinfo {author} {\bibfnamefont {J.}~\bibnamefont {Kelly}}, \bibinfo {author} {\bibfnamefont {Z.}~\bibnamefont {Chen}}, \bibinfo {author} {\bibfnamefont {M.}~\bibnamefont {Neeley}}, \bibinfo {author} {\bibfnamefont {A.}~\bibnamefont {Megrant}}, \bibinfo {author} {\bibfnamefont {B.}~\bibnamefont {Burkett}}, \bibinfo {author} {\bibfnamefont {R.}~\bibnamefont {Barends}}, \bibinfo {author} {\bibfnamefont {K.}~\bibnamefont {Arya}}, \bibinfo {author} {\bibfnamefont {B.}~\bibnamefont {Chiaro}}, \bibinfo {author} {\bibfnamefont {Y.}~\bibnamefont {Chen}}, \bibinfo {author} {\bibfnamefont {A.}~\bibnamefont {Dunsworth}}, \bibinfo {author} {\bibfnamefont {A.}~\bibnamefont {Fowler}}, \bibinfo {author} {\bibfnamefont {B.}~\bibnamefont {Foxen}}, \bibinfo {author} {\bibfnamefont {C.}~\bibnamefont {Gidney}}, \bibinfo {author} {\bibfnamefont {M.}~\bibnamefont {Giustina}}, \bibinfo {author} {\bibfnamefont {R.}~\bibnamefont {Graff}}, \bibinfo
  {author} {\bibfnamefont {T.}~\bibnamefont {Huang}}, \bibinfo {author} {\bibfnamefont {E.}~\bibnamefont {Jeffrey}}, \bibinfo {author} {\bibfnamefont {E.}~\bibnamefont {Lucero}}, \bibinfo {author} {\bibfnamefont {J.}~\bibnamefont {Mutus}}, \bibinfo {author} {\bibfnamefont {O.}~\bibnamefont {Naaman}}, \bibinfo {author} {\bibfnamefont {C.}~\bibnamefont {Neill}}, \bibinfo {author} {\bibfnamefont {C.}~\bibnamefont {Quintana}}, \bibinfo {author} {\bibfnamefont {P.}~\bibnamefont {Roushan}}, \bibinfo {author} {\bibfnamefont {D.}~\bibnamefont {Sank}}, \bibinfo {author} {\bibfnamefont {A.}~\bibnamefont {Vainsencher}}, \bibinfo {author} {\bibfnamefont {J.}~\bibnamefont {Wenner}}, \bibinfo {author} {\bibfnamefont {T.}~\bibnamefont {White}}, \bibinfo {author} {\bibfnamefont {S.}~\bibnamefont {Boixo}}, \bibinfo {author} {\bibfnamefont {R.}~\bibnamefont {Babbush}}, \bibinfo {author} {\bibfnamefont {V.}~\bibnamefont {Smelyanskiy}}, \bibinfo {author} {\bibfnamefont {H.}~\bibnamefont {Neven}},\ and\ \bibinfo {author}
  {\bibfnamefont {J.~M.}\ \bibnamefont {Martinis}},\ }\bibfield  {title} {\bibinfo {title} {Fluctuations of energy-relaxation times in superconducting qubits},\ }\href {https://doi.org/10.1103/PhysRevLett.121.090502} {\bibfield  {journal} {\bibinfo  {journal} {Phys. Rev. Lett.}\ }\textbf {\bibinfo {volume} {121}},\ \bibinfo {pages} {090502} (\bibinfo {year} {2018})}\BibitemShut {NoStop}%
\bibitem [{\citenamefont {Piacente}\ \emph {et~al.}(2004)\citenamefont {Piacente}, \citenamefont {Schweigert}, \citenamefont {Betouras},\ and\ \citenamefont {Peeters}}]{piacente2004}%
  \BibitemOpen
  \bibfield  {author} {\bibinfo {author} {\bibfnamefont {G.}~\bibnamefont {Piacente}}, \bibinfo {author} {\bibfnamefont {I.~V.}\ \bibnamefont {Schweigert}}, \bibinfo {author} {\bibfnamefont {J.~J.}\ \bibnamefont {Betouras}},\ and\ \bibinfo {author} {\bibfnamefont {F.~M.}\ \bibnamefont {Peeters}},\ }\bibfield  {title} {\bibinfo {title} {Generic properties of a quasi-one-dimensional classical {Wigner} crystal},\ }\href {https://doi.org/10.1103/PhysRevB.69.045324} {\bibfield  {journal} {\bibinfo  {journal} {Phys. Rev. B}\ }\textbf {\bibinfo {volume} {69}},\ \bibinfo {pages} {045324} (\bibinfo {year} {2004})}\BibitemShut {NoStop}%
\bibitem [{\citenamefont {Ikegami}\ \emph {et~al.}(2015)\citenamefont {Ikegami}, \citenamefont {Akimoto},\ and\ \citenamefont {Kono}}]{ikegamiMeltingWignerCrystal2015}%
  \BibitemOpen
  \bibfield  {author} {\bibinfo {author} {\bibfnamefont {H.}~\bibnamefont {Ikegami}}, \bibinfo {author} {\bibfnamefont {H.}~\bibnamefont {Akimoto}},\ and\ \bibinfo {author} {\bibfnamefont {K.}~\bibnamefont {Kono}},\ }\bibfield  {title} {\bibinfo {title} {Melting of {Wigner} crystal on helium in quasi-one-dimensional geometry},\ }\href {https://doi.org/10.1007/s10909-014-1272-8} {\bibfield  {journal} {\bibinfo  {journal} {Journal of Low Temperature Physics}\ }\textbf {\bibinfo {volume} {179}},\ \bibinfo {pages} {251} (\bibinfo {year} {2015})}\BibitemShut {NoStop}%
\end{thebibliography}%

\vspace{0.4cm}

\noindent \textbf{\large Acknowledgments}

\noindent We are grateful to M.I.~Dykman, N.O.~Birge, J.R.~Lane, J.M.~Kitzman, G.~Koolstra and K.~Nasyedkin for valuable discussions. We also thank R.~Loloee and D.~Edmunds for technical assistance and B.~Bi for device fabrication advice and use of the W.M.~Keck Microfabrication Facility at MSU. The Michigan State University portion of his work was supported by the National Science Foundation via grant number DMR-2003815 (JP) and the Cowen Family Endowment at MSU.

\newpage

\noindent \textbf{\large Supplementary Information}

\vspace{1cm}

\section{Microchannel plasmon dispersion}

\noindent In the main text we utilize a two-dimensional plasmon dispersion relation (Equation (1)) to study the density dependence of the plasmon resonance frequencies extracted from the experimental data (as shown in Fig.~3 of the main manuscript). While this expression is relevant for the density range over which we observe plasmons in the microchannel device, it is instructive to also discuss the case of a one-dimensional plasmon, which could in principle be realized in the limit of low electron density and tight spatial confinement.

In the low-density limit, the electron system can form a single electron chain due to the constraints imposed by the confining potential of the central microchannel. The problem of the plasmon dispersion for a single linear row of electrons is analytically solvable~\cite{piacente2004} and, for a screened Coulomb interaction, is given by:

\begin{equation}
    \omega_{1}^2(q) = \frac{e^2}{4 \pi \varepsilon_0m_e} \ n_l q^2 \Big[\frac{3}{2} + \ln{\frac{n_l}{\kappa}}  \Big].
    \label{eq:piacente}
\end{equation}

\noindent Here, $q = n \pi /L$ is the plasmon wave number associated with mode number $n$ and channel length $L$, $n_l$ is the linear density of the electron chain, and the parameter $\kappa = 1/\lambda$ is the reciprocal of the screening length $\lambda$. This expression only strictly applies for the case of a single electron chain and we note that in our experiments the number of electron chains increases with increasing channel electrode bias voltage $V_{\mathrm{ch}}$. However, for the sake of comparison, we can assume that at low density (i.e. in the vicinity of the threshold voltage $V_{\mathrm{ch}} \gtrsim V^{\mathrm{th}}_{\mathrm{ch}}$) the plasmons in the microchannel can be described by the excitations of individual electron chains having a dispersion given by Equation~\ref{eq:piacente}. This simplification allows us to compare a quasi-one-dimensional model with our experimental data as well as the two-dimensional plasmon dispersion given by Equation~(1) in the main manuscript. In Supplementary Fig.~\ref{fig:plas_comp}a, we plot Equation~(\ref{eq:piacente}) along with the data and the two-dimensional screened dispersion relation $\omega_p$ from Equation~(1). As shown in the figure, we find that the simplified single electron chain model and the two-dimensional plasmon dispersion tend to converge at low electron density. As expected, we also find that as the density increases, and the system becomes more two-dimensional, the simplified single electron chain model only very roughly agrees with our data. In particular the scaling of the plasmon harmonics and the density dependence are not well-captured by the one-dimensional dispersion. In contrast, the fully two-dimensional plasmon dispersion relation more accurately captures the observed resonances.
\begin{figure}
    \centering
    \includegraphics[width = 0.4\textwidth]{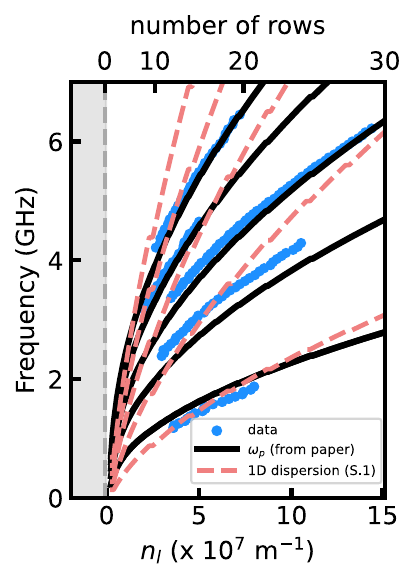}
    \caption{Comparison of the one-dimensional plasmon dispersion relation from Eq.~\ref{eq:piacente} to the dispersion relation used in the main text $\omega_p$ (solid black lines) along with the data (blue dots) for the first five plasmon modes. The linear electron density $n_l$ and the number of electron rows are calculated using finite element modeling (FEM).} 
    \label{fig:plas_comp}
\end{figure}

These results are also consistent with the dimensionality of the electron system inferred from the melting properties of a microchannel confined Wigner solid. The ground state electron configuration of a non-degenerate two-dimensional electron system is determined by the plasma parameter $\Gamma = e^2 \sqrt{\pi n_s}/4 \pi \epsilon_0 k_B T$ -- the ratio of the Coulomb potential energy to the mean kinetic energy. When $\Gamma < \Gamma^{\textrm{2D}}_{\mathrm{c}}=130$, the thermal fluctuations dominate and the electron system behaves like a fluid, while above the critical value of $\Gamma^{\textrm{2D}}_{\mathrm{c}}$, a two-dimensional Wigner solid forms. Previous transport experiments on microchannel confined electron systems~\cite{rees2016structural,ikegamiMeltingWignerCrystal2015} have shown that the melting of the Wigner solid is well-described by a value of $\Gamma^{\textrm{2D}}_{\mathrm{c}}$ for a number of electron rows in the channel $N_y \gtrsim 10-20$, while for small numbers of rows the critical plasma parameter is suppressed. Finally we note that, in our experiments we do not observe plasmons until \vch~$ \simeq 0.45$~V, which corresponds to an electron row number across the channel of $N_y\simeq 10$ according to FEM calculations. For this number of rows the thermodynamic properties of the electrons in the microchannel are consistent with the properties of a two-dimensional electron system.

\section{Non-equilibrium transport in the presence of microwave excitation}
\begin{figure}[b]
    \centering
    \includegraphics[width = 0.7\textwidth]{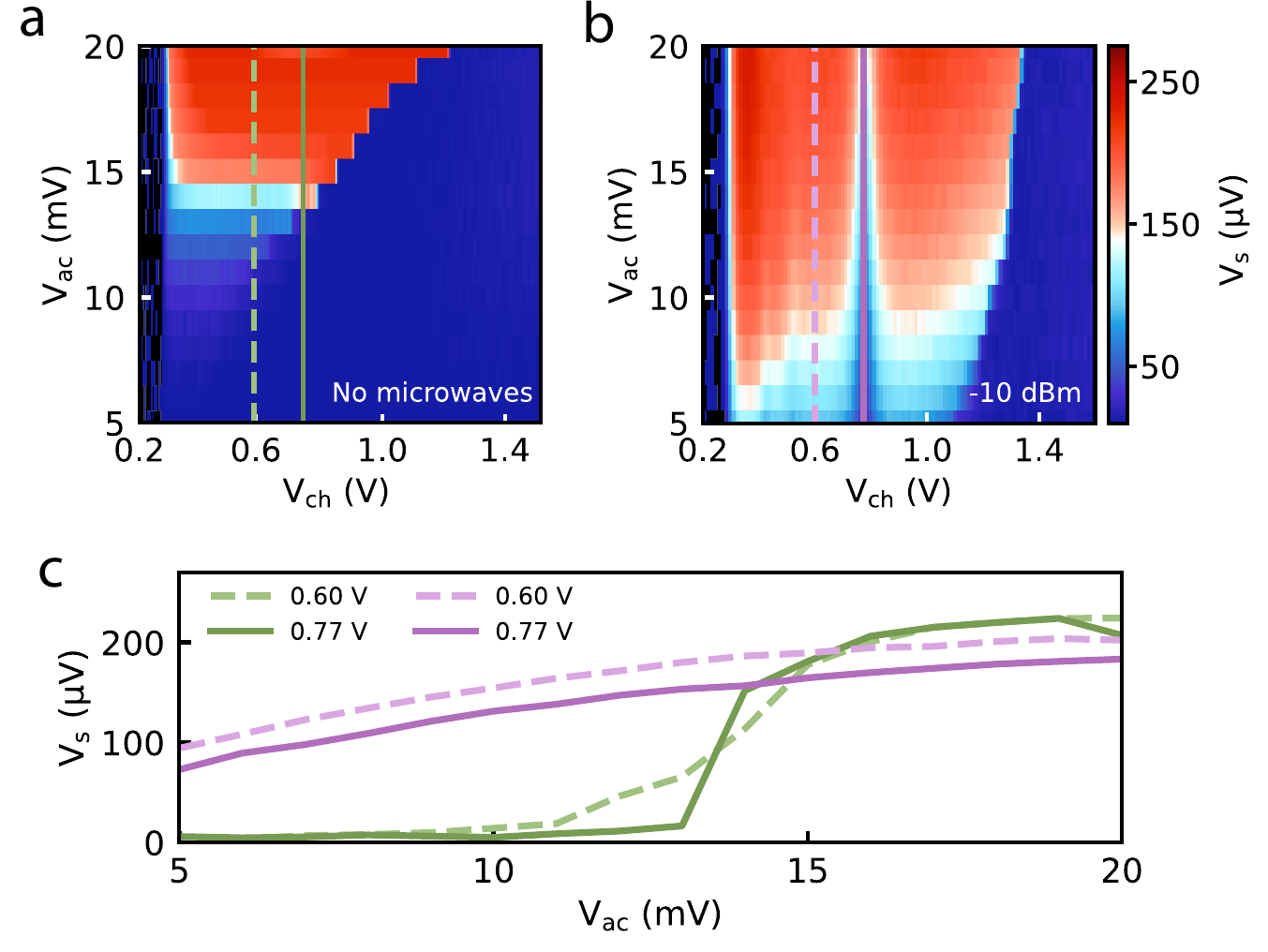}
    \caption{Drive-dependent transport measurements at $T = 16$~mK with and without microwave side gate excitation. (a) Density-dependent microchannel transport with increasing ac drive voltage, \vac, with no microwave power applied to the side gate electrode. This measurement was performed at  {\fac~$=3$~MHz}, {\vres~$=0.7$~V}, and {\vgt~$=0$~V}. Green dashed {(\vch~$=0.6$~V)} and solid {(\vch~$=0.77$~V)} lines correspond to the linecuts plotted in (c). (b) Transport measurements similar to those presented in panel a) except with the additional application of a {$\omega/2\pi = 5.5$~GHz} microwave tone applied to the side gate electrode with $P = -10$~dBm. Purple dashed {(\vch~$=0.6$~V)} and solid {(\vch~$=0.77$~V)} lines correspond to the linecuts plotted in (c). (c) For additional clarity we show vertical line cuts of the transport signal from panel (a) (green dashed and solid) and panel~(b)~(purple dashed and solid).}
    \label{fig:vac}
\end{figure}
To characterize the non-equilibrium transport response of the electron system in the presence of the microwave excitation field, we perform a series of ac drive dependent transport measurements with and without simultaneous microwave power applied to the side gate electrodes. Supplementary Fig.~\ref{fig:vac}a shows the density dependent transport through the central microchannel with increasing ac drive amplitude and with no microwave excitation on the side gate electrode. At low ac drive (\vac~$< 12$~mV), the electron system remains in a low-conductivity Wigner solid state for all values of the electron density shown in Supplementary Fig.~\ref{fig:vac}a. In this regime, the electron velocity saturates at the ripplon velocity as described in the main text. At \vac~$\simeq 12$~mV, the electron system conductivity abruptly increases, indicative of the transition into a high-conductivity state of the electron system, which can be interpreted as either an unpinned Wigner solid or a disordered electron liquid~\cite{saitoh1977warm, shirahama1995dynamical, ikegami2009nonlinear, ikegamiMeltingWignerCrystal2015, rees2016stick, Dykman2016Phys, rees2020dynamical}.

In Supplementary Fig.~\ref{fig:vac}b we present a similar ac drive dependent transport measurement but with a $\omega/2\pi = 5.5$~GHz microwave tone applied to the side gate electrode. The presence of the $n=3$ plasmon mode appears as a reduction in the measured transport signal (solid purple line) as described in the main text. Additionally, and in contrast to the case with no microwave excitation (Supplementary Fig.~\ref{fig:vac}a), the non-linear Bragg-Cherenkov regime is not observed down to the lowest levels of ac drive for which we are able to perform transport measurements. Rather we observe steady increase in the transport signal with increasing ac drive amplitude both on and off resonance (see purple linecuts in Supplementary Fig.~\ref{fig:vac}c), which is consistent with the heating of an underlying electron liquid state.

\section{Plasmon mode generation and coupling}
\begin{figure}
    \centering
    \includegraphics[width = 0.8\textwidth]{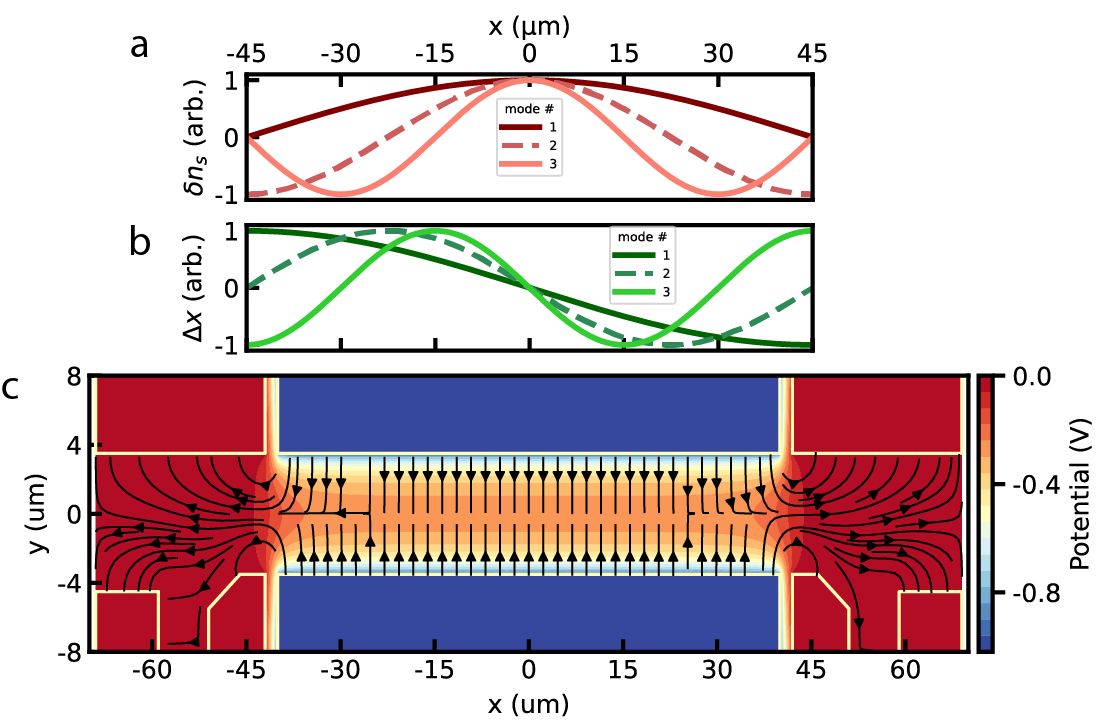}
    \caption{Plot of how the first three plasmon modes will manifest (a) in terms of the change in areal electron density $\delta n_s$ along the channel and (b) in terms of the change in electron displacement $\Delta x$ along the channel. The odd modes ($n = 1$ and $n = 3$) are represented by solid lines and the even mode ($n = 2$) by a dashed line. (c) Plot of the microwave potential in the region of the central microchannel with corresponding electric field lines (black lines with arrows) calculated from FEM simulations. The plot shows how the field lines terminate on the various areas of the device, promoting a stronger coupling to odd versus even plasmon modes.}
    \label{fig:stream}
\end{figure}

\noindent In the experiments, plasmons are generated via a microwave excitation signal applied to the gate electrode (colored blue in Supplementary Fig.~\ref{fig:stream}c). The coupling between the microwaves and plasmons in the central channel is determined by the dot product of the electric field distribution generated by top gate electrodes and electron displacement field $\int(\vec{E} \cdot \vec{\Delta x}) dxdy$. In Supplementary Fig.~\ref{fig:stream} we schematically plot the spatial distribution of the charge density displacement $\delta n_s$ (Supplementary Fig.~\ref{fig:stream}a) and corresponding electron position displacement $\Delta x$ (Supplementary Fig.~\ref{fig:stream}b) for the first three longitudinal plasmon modes along the channel, as well as the electric field distribution calculated using FEM (Supplementary Fig.~\ref{fig:stream}c). For longitudinal plasmon standing-waves along the channel length, the primary contribution to the coupling comes from the regions near the channel ends, where the $x$-component of the microwave electric field is nonzero. The symmetry of the electric field also determines which modes couple more efficiently: modes with position displacement values of opposite sign at the channel ends — corresponding to odd-numbered modes — are more readily excited by microwave fields applied to the gate electrode. We also note that, in principle, the electric field configuration within the channel allows for the excitation of transverse modes. These plasmons correspond to the optical branch of the plasmon excitation spectrum, with frequency increasing as the wavevector increases and a finite frequency gap in the limit of small wavevector, which is set by transverse confining potential frequency $\omega_0$ (for number of rows $N_y \geq 2$)~\cite{piacente2004}. We estimate $\omega_0/2\pi > 10$~GHz for the voltage ranges used in our experiments, making the excitation of transversal modes unlikely.

\section{Fitting of plasmon resonances}

To investigate the power dependence of the plasmon resonances presented in the Fig.~4 of the main text, we fit each transport trace to a Lorentzian distribution and a Gaussian distribution, where we have taken into account the background transport signal with the addition of a logistic function, which we find phenomenologically captures the smoothly varying density-dependent background. The composite fitting function containing the Lorentzian distribution has the following form,

\begin{equation}
    \mathcal{L}(n_s) = \frac{\gamma}{\pi [(n_s - n_{s0})^2 + \gamma^2]} + A \Big[ 1 - \frac{1}{1 - e^{(n_s - n_{s1})/\alpha}} \Big],
    \label{eq:lor_SI}
\end{equation}

\noindent where $2 \gamma$ is the full-width at half-maximum of the Lorentzian, $n_{s0}$ is the density at which the resonance feature is centered and \{$A$, $n_{s1}$, $\alpha$\} are phenomenological fitting parameters describing the logistic function. The electron densities are mapped from the corresponding values of $V_{\mathrm{ch}}$ in the transport data, calculated using Equation~(3) from the main text. We find that the low-power linecuts are well-fit using this function and an example of the fit is shown in Supplementary Fig.~\ref{fig:lineshape}a.

The composite function containing the Gaussian distribution, which we find fits the high power data better, has the form
\begin{equation}
    \mathcal{G}(n_s) =  \frac{e^{-(n_s - n_{s0})^2/2\sigma^2}}{\sigma \sqrt{2 \pi}} + A \Big[ 1 - \frac{1}{1 - e^{(n_s - n_{s1})/\alpha}} \Big].
    \label{eq:gau_SI}
\end{equation}
\noindent where $2 \sigma$ is the linewidth of Gaussian distribution and the second term describes background signal in a similar way as in Eq.\ref{eq:lor_SI}. An example of the fit to the higher power data using this function is shown in Supplementary Fig.~\ref{fig:lineshape}b.

\begin{figure}
    \centering
    \includegraphics[width = 1\textwidth]{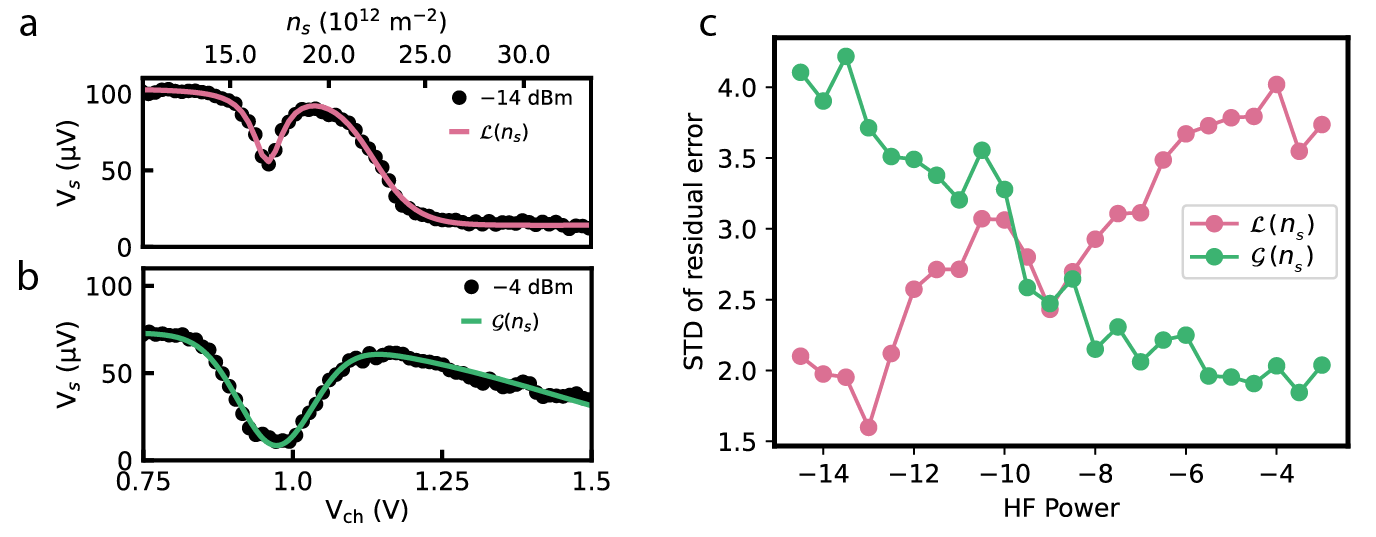}
    \caption{(a),(b) Fit of transport data from Fig.~4b (over an extended range) using the composite functions Eq.~\ref{eq:lor_SI} and Eq.~\ref{eq:gau_SI}. (c) Standard deviation of the residual error from fitting each transport linecut from $P = -14.5$~dBm to $P = -3$~dBm of Fig.~4a of the main text to $\mathcal{G}(n_s)$ (green dots) and to $\mathcal{L}(n_s)$ (pink dots).}
    \label{fig:lineshape}
\end{figure}

In Supplementary Fig.~\ref{fig:lineshape}c, we show how the standard deviation of the residual error evolves when fitting all the microwave power-dependent transport data (from the main text Fig.~4a) to $\mathcal{L}(n_s)$ (pink) and to $\mathcal{G}(n_s)$ (green). We find that plasmon resonance features measured at microwave powers $P < -10$~dBm are better represented with the Lorentzian-like function, while at higher powers, the Gaussian-like lineshape provides a better description of the data.

\end{document}